\pdfoutput=1

\documentclass[preprint,5p]{elsarticle}

\usepackage{graphicx,amssymb}
\usepackage{epsf,psfig}
\usepackage{txfonts}
\usepackage{multirow}
\usepackage{bigstrut}

\journal{New Astronomy}

\begin{document}

\begin{frontmatter}

\title{Exploring the nature of orbits in a galactic model with a massive nucleus}

\author{Euaggelos E. Zotos\corref{}}

\address{Department of Physics, \\
Section of Astrophysics, Astronomy and Mechanics, \\
Aristotle University of Thessaloniki \\
GR-541 24, Thessaloniki, Greece}

\cortext[]{Corresponding author: \\
\textit{E-mail address}: evzotos@astro.auth.gr (Euaggelos E. Zotos)}

\begin{abstract}

In the present article, we use an axially symmetric galactic gravitational model with a disk-halo and a spherical nucleus, in order to investigate the transition from regular to chaotic motion for stars moving in the meridian $(r,z)$ plane. We study in detail the transition from regular to chaotic motion, in two different cases: the time independent model and the time evolving model. In both cases, we explored all the available range regarding the values of the main involved parameters of the dynamical system. In the time dependent model, we follow the evolution of orbits as the galaxy develops a dense and massive nucleus in its core, as mass is transported exponentially from the disk to the galactic center. We apply the classical method of the Poincar\'{e} $(r,p_r)$ phase plane, in order to distinguish between ordered and chaotic motion. The Lyapunov Characteristic Exponent is used, to make an estimation of the degree of chaos in our galactic model and also to help us to study the time dependent model. In addition, we construct some numerical diagrams in which we present the correlations between the main parameters of our galactic model. Our numerical calculations indicate, that stars with values of angular momentum $L_z$ less than or equal to a critical value $L_{zc}$, moving near to the galactic plane, are scattered to the halo upon encountering the nuclear region and subsequently display chaotic motion. A linear relationship exists between the critical value of the angular momentum $L_{zc}$ and the mass of the nucleus $M_n$. Furthermore, the extent of the chaotic region increases as the value of the mass of the nucleus increases. Moreover, our simulations indicate that the degree of chaos increases linearly, as the mass of the nucleus increases. A comparison is made between the critical value $L_{zc}$ and the circular angular momentum $L_{z0}$ at different distances from the galactic center. In the time dependent model, there are orbits that change their orbital character from regular to chaotic and vise versa and also orbits that maintain their character during the galactic evolution. These results strongly indicate that the ordered or chaotic nature of orbits, depends on the presence of massive objects in the galactic cores of the galaxies. Our results suggest, that for disk galaxies with massive and prominent nuclei, the low angular momentum stars in the associated central regions of the galaxy, must be in predominantly chaotic orbits. Some theoretical arguments to support the numerically derived outcomes are presented. Comparison with similar previous works is also made.

\end{abstract}

\begin{keyword}
Galaxies: kinematics and dynamics;
\end{keyword}

\end{frontmatter}

\section{Introduction}

Central mass concentrations (CMCs) are frequently found in galaxies of all types, such as barred, spiral, disk or elliptical galaxies. A few examples of CMCs are:

1. Large condensations of molecular gas with scales of 0.1 $\sim$ 2kpc and masses of $10^{7} \sim 10^{9}$ $M_{\odot}$ are found in the central regions (e.g., Ohta, et al., 1996; Sakamoto et al., 1999; Regan et al., 2001). They are particularly evident in barred galaxies, where they are believed to be created by bar-driven inflow (e.g., Athanassoula, 1992; Heller \& Shlosman, 1994). The contribution of gas in the galactic evolution is a very important parameter. The reader can find interesting information about the role of the gas in the review of Kormendy \& Kennicutt (2004) and references therein.

2. Today it is widely accepted that most, if not all, of the galaxies have in their cores super-massive black holes (SMBHs). Central supermassive black holes seem to be a ubiquitous component in spiral disk galaxies, as well as in elliptical galaxies. Typical masses are in the range of $10^{6} \sim 10^{9}$ $M_{\odot}$ (e.g., Young, 1980, Turner, 1991; Magorrian et al., 1998; Ferrarese \& Merritt, 2000; Gebhardt et al., 2000; McLure \& Dunlop, 2001; Tremaine et al., 2002; Wada \& Norman, 2002; Kawakatsu \& Umemura, 2002; Hopkins, et al., 2006; Shen, et al., 2008). The reader can find more details regarding SMBHs in the review of Kormendy \& Richstone, 1995.

3. Dense star clusters are found near the centers of many spiral galaxies (Carollo, 2004; Walcher et al., 2003); these are
young, very compact (a few to up to 20 pc), and relatively massive ($10^{6} \sim 10^{8}$ $M_{\odot}$).

For the purposes of this study, a CMC is a sufficiently large mass at the center of a galaxy that is likely to have a dynamical effect on the evolution of its host galaxy. In other words, we focus on the dynamical consequences of a dense and massive nucleus.

Dynamic searches indicate the presence of massive objects in the central regions of most galaxies. (e.g., M31, M87 and NGC 4594). The estimation regarding the mass of the central region range from $10^{6} M_{\odot}$ up to about $10^{10} M_{\odot}$. Improved ground-based instruments and especially the Hubble Space Telescope (H.S.T), have given interesting information about the central regions of the galaxies. Dynamical evidence for the presence of a massive object in the core of a galaxy, is based on the value of the mass-to-luminosity ratio $M/L$. If this ratio increases towards the galactic center to values that are several times larger than normal, then this strongly indicates that a massive object has been discovered (Kormendy \& Rihstone, 1995). Rapid rotation is also a strong indication for the presence of a massive object in the core of a galaxy (Miyoshi et al., 1995; Greenhill et al., 1997; Rubin et al., 1997; Bertola et al., 1998; Sofue et al., 1999).

The effect of a central mass concentration (CMC) on the phase space in global galactic models has already been studied extensively in several earlier research works. Hasan \& Norman (1990) have shown that a black hole or a CMC in a barred galaxy can dissolve the barred structure. Moreover, the fundamental orbits of the system (called B orbits), supporting the bar are orbits elongated in the direction of the bar. As the black hole's mass increases, an inner Lindblad resonance (ILR) appears and moves outward. Thus, the B orbits disappear when the ILR reaches the end of the bar. This effect is significantly increased, if the bar is thinner. Orbits that stay closer to the central mass, with smaller Jacobi constants, can change their character from regular to chaotic more easily. Furthermore, the study of Hasan et al. (1993) indicates that, as the mass of the CMC increases the stable regular orbits in the region where the bar potential competes in strength with the central mass potential, especially around the region of the ILR will be no longer present and their position will be occupied by chaotic orbits. Shen \& Sellwood (2004) have conducted a systematic study of the effects of CMC on bars, using high quality $N$-body simulations. They have experimented with both strong and weak initial bars and a wide range of the physical parameters of the CMC, such as the final mass, the scale length and the mass growth time. Their outcomes suggest that, for a given mass compact CMCs (such as super-massive black holes) are more destructive to barred structures than are more diffuse ones (such as molecular gas clouds in many galactic centers). They have shown that the former are more efficient scatters of bars supporting regulars elongated orbits, that pass close to the galactic center and therefore, decrease the percentage of the the regular orbits and increase the area of the chaotic region in the phase space. All the above outcomes strongly indicate that a massive nucleus in the central parts of a galaxy, is a very important parameter of the dynamical system.

In several previous works, we see that low angular momentum stars in galaxies with massive nuclei, are scattered to much higher scale heights, displaying chaotic motion (see Caranicolas, 1997, Karanis \& Caranicolas, 2001; Caranicolas \& Zotos, 2010; Zotos, 2011a,b). All the above studies, not only describe the chaotic character of orbits of low angular momentum stars, but also provide the main reason which is responsible for this behavior. The reason is the presence of the strong   force, which comes from the massive nucleus which resides in the core of the galaxy. The dynamical models, used in all previous papers were global models. In Caranicolas (1997), it was used the mass model introduced by Carlberg \& Innanen (1987), while in Karanis \& Caranicolas (2001) they have used a logarithmic potential. The most characteristic result, derived in all cases, was the linear dependence of the critical value of angular momentum $L_{zc}$ and the mass of the nucleus $M_n$. From our point of view, the background of this series of paper, in combination of other orbit calculations, was that managed to find and present, both numerically and semi-theoretically, relationships connecting the involved physical parameters of the dynamical system, such as the angular momentum and the degree of chaos in disk galaxies.

Recently, Caranicolas \& Zotos (2011) used a model consisting of two components. The first component was a logarithmic potential describing a prolate elliptical galaxy, while the second one was a potential of a dense spherical nucleus. Two different cases were studied. The case when the nucleus is absent and the case were we have an active galaxy hosting a dense and massive nucleus. The main outcome of this work is that low angular momentum stars on approaching a dense and massive nucleus are deflected to higher values of $z$ displaying chaotic motion. This suggests that it is the strong vertical force near the dense nucleus that is the main responsible for this scattering, combined with the star's low angular momentum which allows the star to approach sufficiently enough to the dense nucleus in order to be deflected.

In an earlier work, Caranicolas \& Papadopoulos (2003) adopted a simple time dependent model in order to study the transition from regularity to chaos in a galaxy, when mass is transported exponentially from the disk to the galactic core forming a dense and massive nucleus. The results of this research suggest that during the galactic evolution, low angular momentum stars change their orbital character from regular to chaotic. In the present work, we will continue this study in more detail. In particular, we shall deploy this gravitational galactic model, in order to investigate the change of the orbital nature for stars moving in the meridian $(r,z)$ plane. We will prove using a systematic and detailed study, that stars do not only change their nature from regular to chaotic during the galactic evolution, but there are also stars which maintain their character and also a small portion of stars that alter their character from chaotic to regular.

The purpose of this research work is: (1) to find the correlations between the involved parameters of the system, (2) to explain the numerically obtained results using theoretical arguments, (3) to compare the critical value of the angular momentum $L_{zc}$ with the value of the circular angular momentum $L_{z0}$ at different distances from the center of the galaxy and (4) to study the evolution of orbits when mass is transported exponentially from the disk to the nucleus.

The present article is organized as follows: In Section 2 we present our gravitational dynamical model. In Section 3 we provide a detailed analysis of the structure of our time independent dynamical system. Moreover, some interesting numerical results are also provided in the same Section. In Section 4 we present some semi-theoretical arguments, in order to support and explain our numerically obtained outcomes. In the next Section we follow the evolution of the orbits in the time dependent model, when mass is transported from the disk to the nucleus. We conclude with Section 6, where the discussion and the conclusions of this research are presented.

\section{The dynamical model}

A very convenient method in order to study the behavior of the orbits of the stars in a galaxy, is to represent the galactic system by using a dynamical model, having about the same dynamical properties as the galaxy. In the present article, we shall study the transition from regular to chaotic motion, in an axially symmetric galaxy described by the disk-halo potential
\begin{equation}
\Phi_d(r,z) = - \frac{M_d}{R_d} \ \ \ ,
\end{equation}
with
\begin{equation}
R_d^2 = b^2 + r^2 + \left[\alpha + \displaystyle\sum\limits_{i=1}^3 \beta_i \sqrt{h_i^2 + z^2}\right]^2 \ \ \ ,
\end{equation}
where $(r,z)$ are the usual cylindrical coordinates. Here $M_d$ is the mass of the disk, $\alpha$ is the scale length of the disk, $h$ corresponds to the disk scale height, while $b$ is the core radius scale length of the halo component. The three components of the disk $\left(\beta_1, \beta_2, \beta_3\right)$, represent the fractional portions of old disk (Population II stellar objects), dark matter and young disk (Population I stellar objects), respectively (see Clutton-Brock et al., 1977; Carlberg \& Innanen, 1987; Caranicolas \& Innanen, 1991). The symbol $\Sigma$ denotes summation over these three components and also should apply that $\beta_1 + \beta_2 +\beta_3 = 1$. At this point,we should explain the role of the dark matter halo, since it is the dominant mass at large radii, where stars are deflected to larger apocenter upon interacting with the dark halo component. The $\beta_2$ component represents the dark matter. In the present research for all our orbital calculations the distance from the galactic center does not exceeds 10kpc. Therefore, we do not study the nature of orbits at large radii at which the dark matter is the dominant mass.

To potential (1) we add a spherically symmetric nucleus
\begin{equation}
\Phi_n(r,z) = - \frac{M_n}{\sqrt{r^2 + z^2 + c_n^2}} \ \ \ ,
\end{equation}
where $M_n$ is the mass and $c_n$ represents the scale length of the nucleus. The Plummer sphere we use in order to increase the central mass, has been applied many times in the past, in order to study the effect of the introduction of a central mass component in a galaxy (see Hasan and Norman, 1990; Hasan et al., 1993).

Thus, the total potential is
\begin{equation}
\Phi_{tot}(r,z) = \Phi_d(r,z) + \Phi_n(r,z) \ \ \ .
\end{equation}

In this research, we use a system of galactic units, where the unit of mass is $2.325 \times 10^7 M_\odot$, the unit of length is 1kpc and the unit of time is $0.997748 \times 10^8$ yr. The unit of the angular momentum is 10 km kpc/s. The velocity unit and the energy unit (per unit mass) are 10 km/s and 100 (km/s)$^2$ respectively, while $G$ is equal to unity. In these units, we use the values: $M_d = 12000$, $\alpha = 3$ (kpc), $b = 8$ (kpc), $\left(\beta_1, \beta_2, \beta_3\right) = (0.4, 0.5, 0.1)$ and $\left(h_1, h_2, h_3\right) = (0.325, 0.090, 0.125)$ kpc, while the mass and the scale length of the nucleus are treated as parameters. The above numerical values of the constant dynamical quantities of the system, secure positive density everywhere and free of singularities.
\begin{figure}[!tH]
\centering
\resizebox{0.95\hsize}{!}{\rotatebox{0}{\includegraphics*{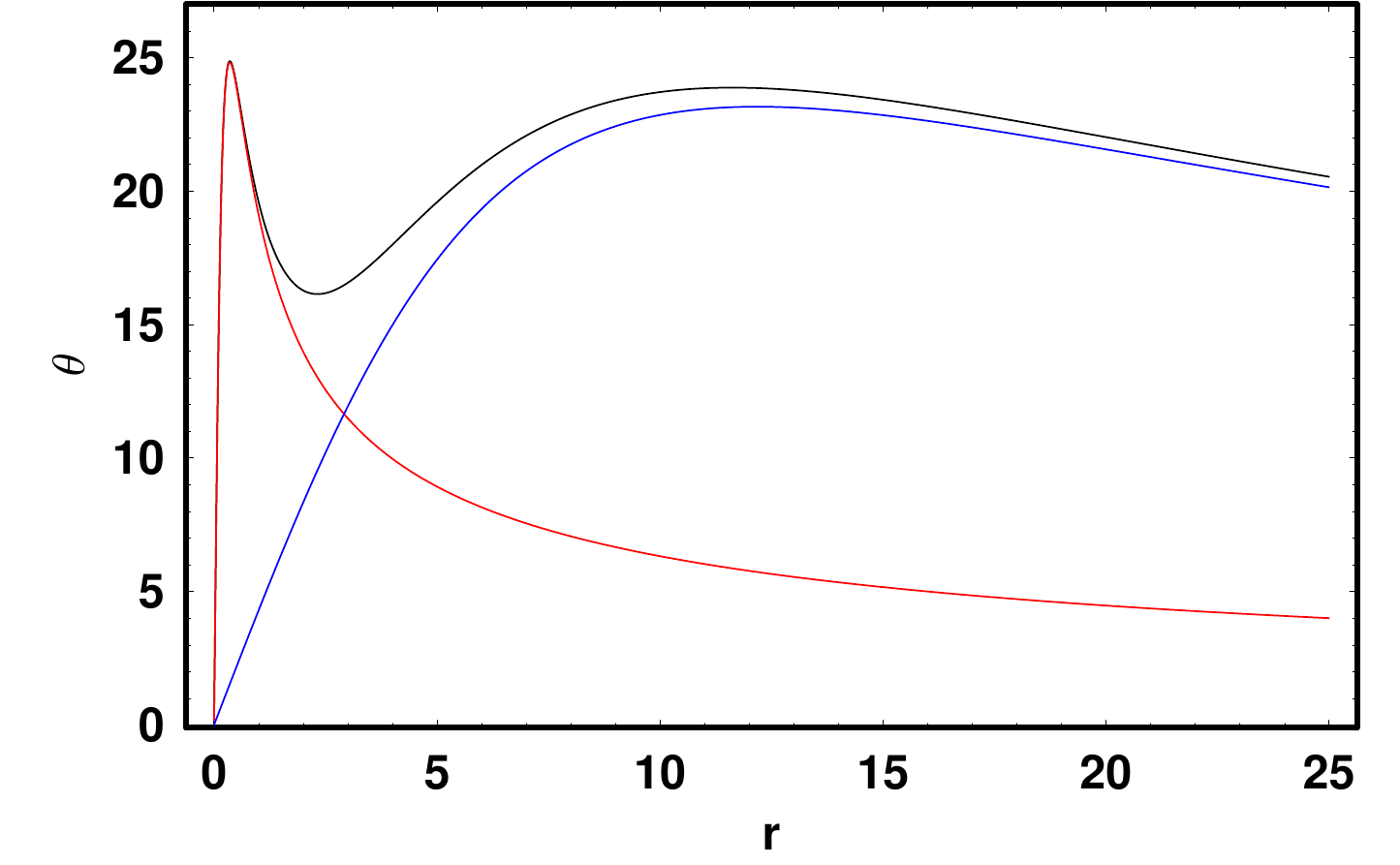}}}
\caption{The rotation curve of the galactic model is shown as the black line. The red line is the contribution from the spherical nucleus, while the blue line is the contribution from the disk-halo potential.}
\end{figure}
\begin{figure*}[!tH]
\centering
\resizebox{0.9\hsize}{!}{\rotatebox{0}{\includegraphics*{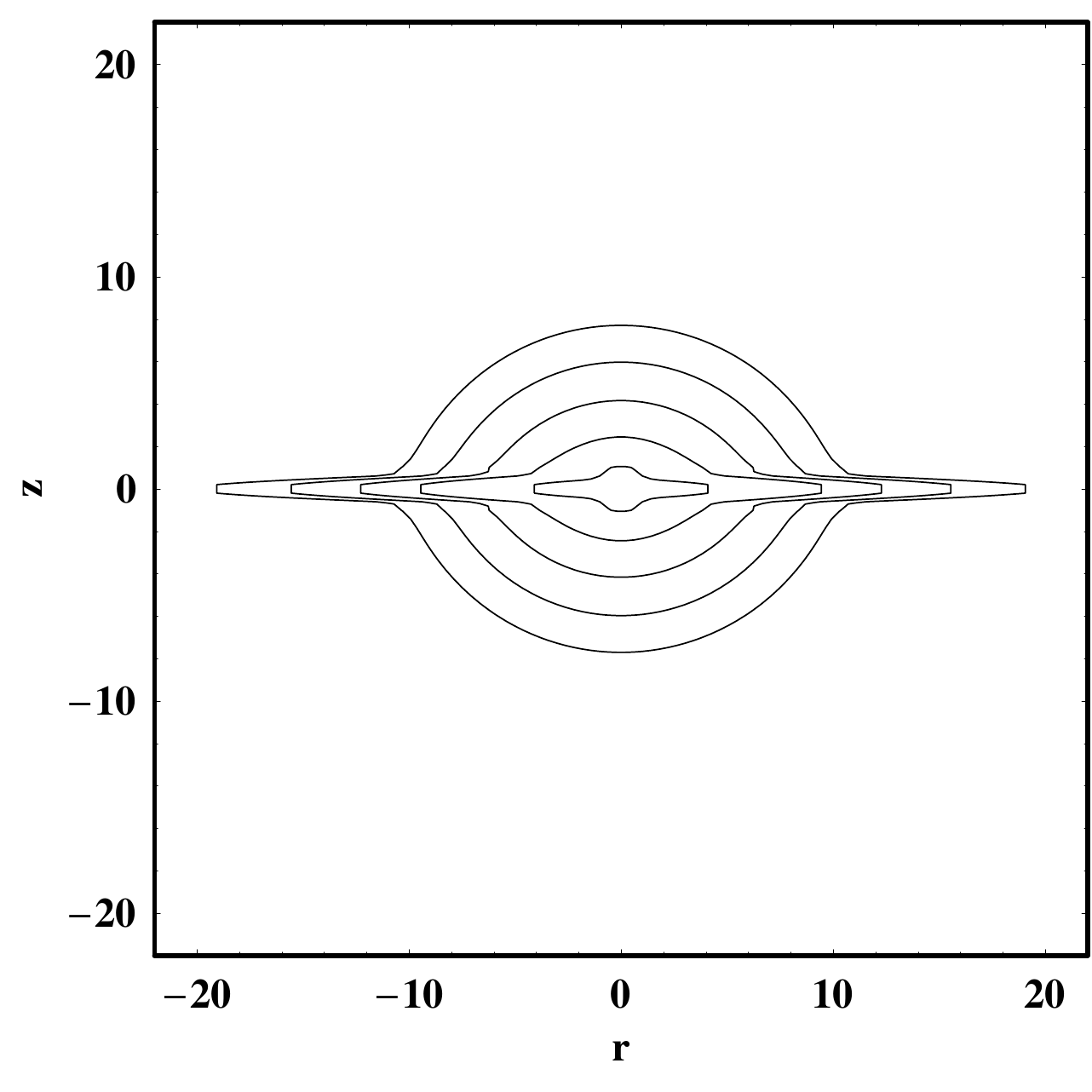}}\hspace{1cm}
                         \rotatebox{0}{\includegraphics*{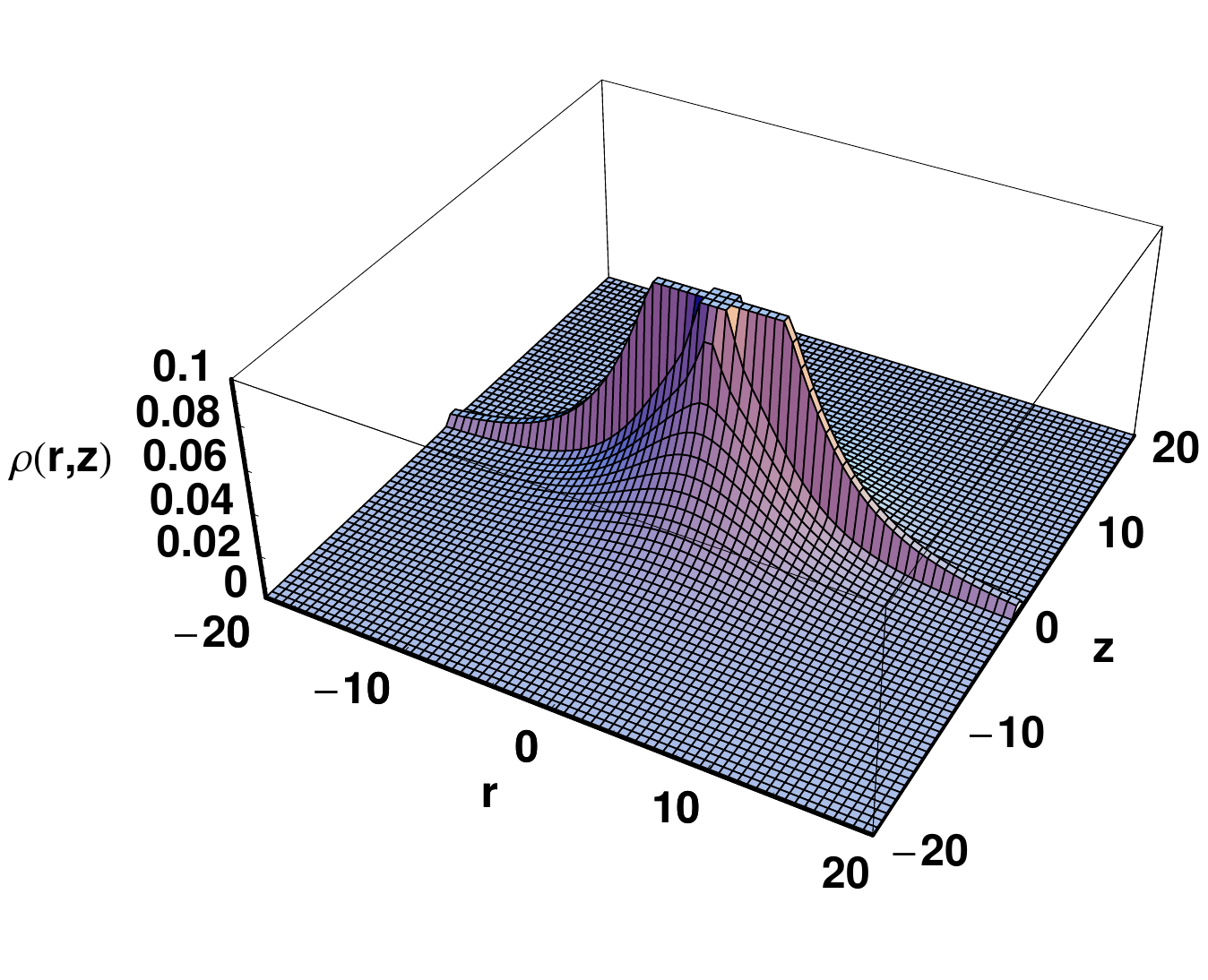}}}
\vskip 0.1cm
\caption{(a-b): (a-left): Density contours for our galactic model and (b-right): A 3D plot of the value of the mass density $\rho$ on the $r - z$ plane. The mass of the nucleus is $M_n=400$, while $c_n=0.25$.}
\end{figure*}

The rotation curve of the galactic model when $M_n=400$ and $c_n=0.25$, is shown as the black line in Figure 1. In the same plot, the red line is the contribution from the spherical nucleus, while the blue line is the contribution from the disk-halo component. We observe, that at small distances from the galactic center $r\leq 2$ kpc, dominates the contribution from the spherical nucleus, while at larger distances $r > 2$ kpc the disk-halo contribution is the dominant factor.

It would be very useful to compute the galactic mass density $\rho(r,z)$ derived from the total potential (4), using the Poisson's equation
\begin{eqnarray}
\rho(r,z) = \frac{1}{4 \pi G} \nabla^2 \Phi_{tot}(r,z) = \nonumber \\
= \frac{1}{4 \pi G} \left[\frac{\partial^2}{\partial r^2} + \frac{1}{r}\frac{\partial}{\partial r}
+ \frac{1}{r^2} \frac{\partial^2}{\partial \phi^2} + \frac{\partial^2}{\partial z^2}\right] \Phi_{tot}(r,z) \ \ \ .
\end{eqnarray}
Remember that due to the axial symmetry the third term in Eq. (5) is zero. Figure 2a shows the contours $\rho(r,z)=const$, for our model. The values of the contours are: (0.010, 0.017, 0.030, 0.052, 0.150). In Figure 2b, we see a 3D plot of the values of the density $\rho(r,z)$ on the $r - z$ plane. We can observe the spherical nucleus at the central region of the galaxy and also the well formed disk structure. This plot assures that the mass density has positive value everywhere. For large values of $r$ and $z$ the mass density varies like $1/r^5$ and $1/z^5$ respectively. This means that the total mass $M(R)$, enclosed in a sphere of radius $R$ increases linearly with distance. This explains why circular velocity profiles, shown in Fig. 1 remains almost flat for large values of $r$. Here we must point out, that our gravitational potential is truncated at $R_{max}=25$ kpc, otherwise the total mass of the galaxy modeled by this potential becomes infinite, which is obviously not physical (see Binney \& Tremaine, 2008).

As the total potential $\Phi_{tot}\left(r,z\right)$ is axially symmetric and the $L_z$ component of the angular momentum is conserved, the dynamical structure of the system, can be studied using the effective potential
\begin{equation}
\Phi_{eff}\left(r,z\right) = \frac{L_z^2}{2r^2} + \Phi_{tot}\left(r,z\right) \ \ \ ,
\end{equation}
in order to study the motion in the meridian $(r,z)$ plane. The equations of motion are
\begin{eqnarray}
\dot{r}=p_r, \ \ \ \dot{z}=p_z, \nonumber \\
\dot{p_r}=-\frac{\partial \ \Phi_{eff}}{\partial r}, \ \ \
\dot{p_z}=-\frac{\partial \ \Phi_{eff}}{\partial z} \ \ \ ,
\end{eqnarray}
where the dot indicates derivative with respect to the time.

The corresponding Hamiltonian can be written as
\begin{equation}
H=\frac{1}{2} \left(p_r^2 + p_z^2 \right) + \Phi_{eff}\left(r,z\right) = E \ \ \ ,
\end{equation}
where $p_r$ and $p_z$, are the momenta per unit mass, conjugate to $r$ and $z$ respectively, while $E$ is the numerical value of the Hamiltonian, which is conserved. Equation (8) is an integral of motion, which indicates that the total energy of the test particle is conserved. The Hamiltonian (8), also describes the motion in the $(r,z)$ plane, rotating at the angular velocity
\begin{equation}
\dot{\phi} = \omega = \frac{L_z}{r^2} \ \ \ .
\end{equation}

All the outcomes of the present work, are based on the numerical integration of the equations of motion (7), which was made by means of a Bulirsh-St\"{o}er method in Fortran 95, with double precision in all subroutines. The accuracy of the calculations was checked by the consistency of the energy integral (8), which was conserved up to the fifteenth significant figure.
\begin{figure}[!tH]
\centering
\resizebox{0.95\hsize}{!}{\rotatebox{0}{\includegraphics*{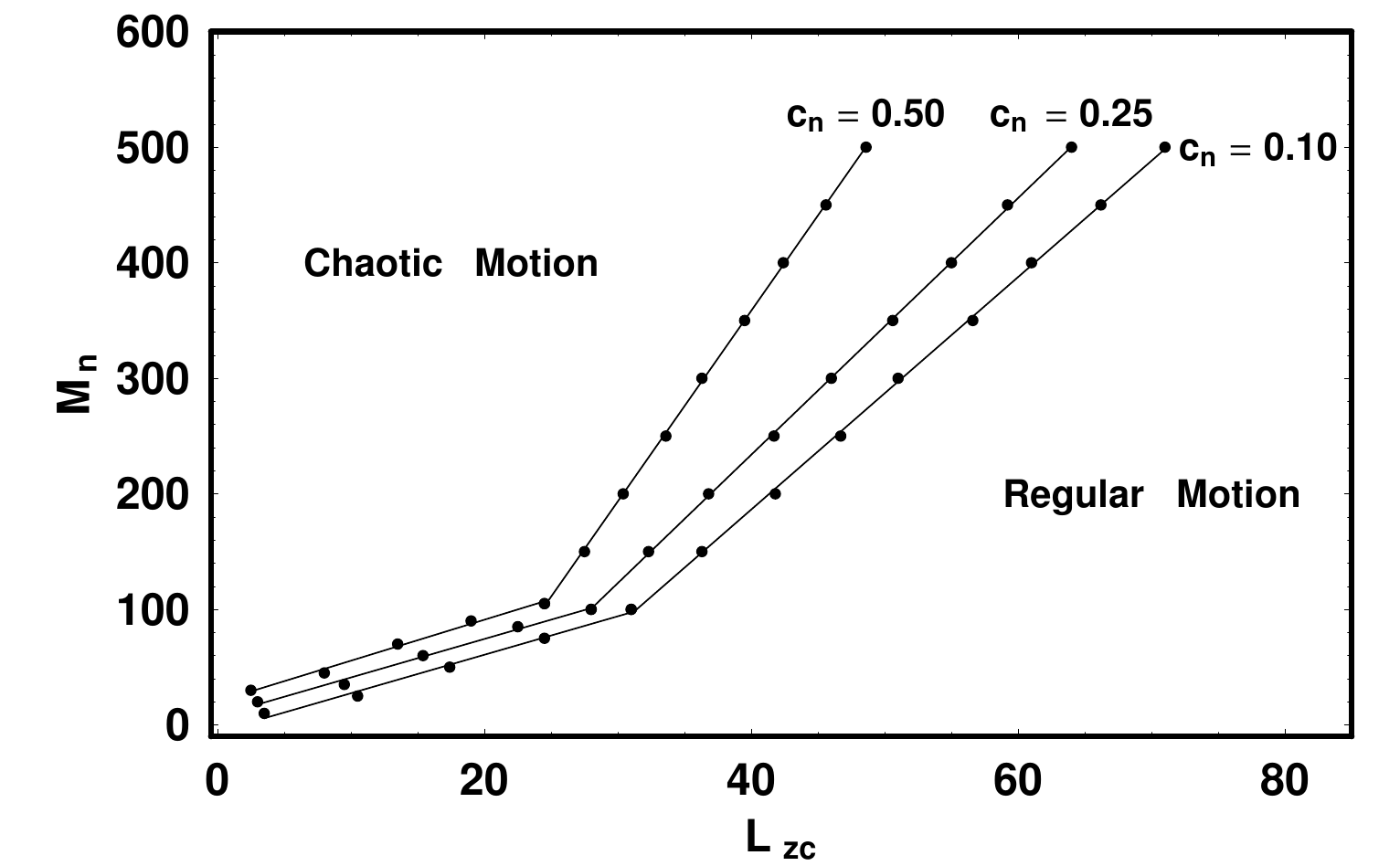}}}
\caption{Relationship between the critical value of the angular momentum $L_{zc}$ and the mass of the nucleus $M_n$. Details are given in the text.}
\end{figure}

\section{The structure of the dynamical system}

In this Section, we shall investigate the regular or chaotic nature of the orbits in our dynamical system. Figure 3 shows the numerically found relationship between the critical value of the angular momentum $L_{zc}$ and the mass of the nucleus $M_n$, for three different values of the scale length of the nucleus $c_n$. The critical value of the angular momentum $L_{zc}$, is the maximum value of the angular momentum $L_z$, for which, stars scattered to the halo display chaotic motion for a given value of the mass of the nucleus $M_n$ and the scale length $c_n$, when all other parameters are kept constant. Orbits were started near $r_0=r_{max}$, with $z_0=p_{r0}=0$, while the initial value of the $p_{z0}$ is always found from the energy integral (8). The value of $r_{max}$ is the maximal root of the equation
\begin{equation}
\frac{L_z^2}{2r^2} + \Phi_{tot} \left(r, z=0\right) = E \ \ \ ,
\end{equation}
which was obtained numerically. The value of the energy integral is $E=-950$ and remains constant during this section. The particular value of the energy was chosen so that in all cases $r_{max} \simeq 10$ and also in order to keep the initial radial and vertical velocities less than 55 km/s. Our numerical experiments indicate that the relationship between $L_{zc}$ and $M_n$ is nearly a straight line in every case (different values of $c_n$). Each line divides the $\left[L_{zc} - M_n\right]$ plane in two parts. Orbits with initial values of the parameters $L_z$ and $M_n$ on the upper left hand side part of the diagram, including the line, are chaotic orbits, while orbits with starting conditions on the lower right part of the same diagram are regular orbits. Note that, for small values of the the angular momentum, the slope of the line is different. Moreover, it is interesting to point out that the extent of the area regarding chaotic motion, increases as the value of $c_n$ becomes smaller, that is when we have a more dense nucleus. We shall come to that point later in the next section.
\begin{figure*}[!tH]
\centering
\resizebox{0.8\hsize}{!}{\rotatebox{270}{\includegraphics*{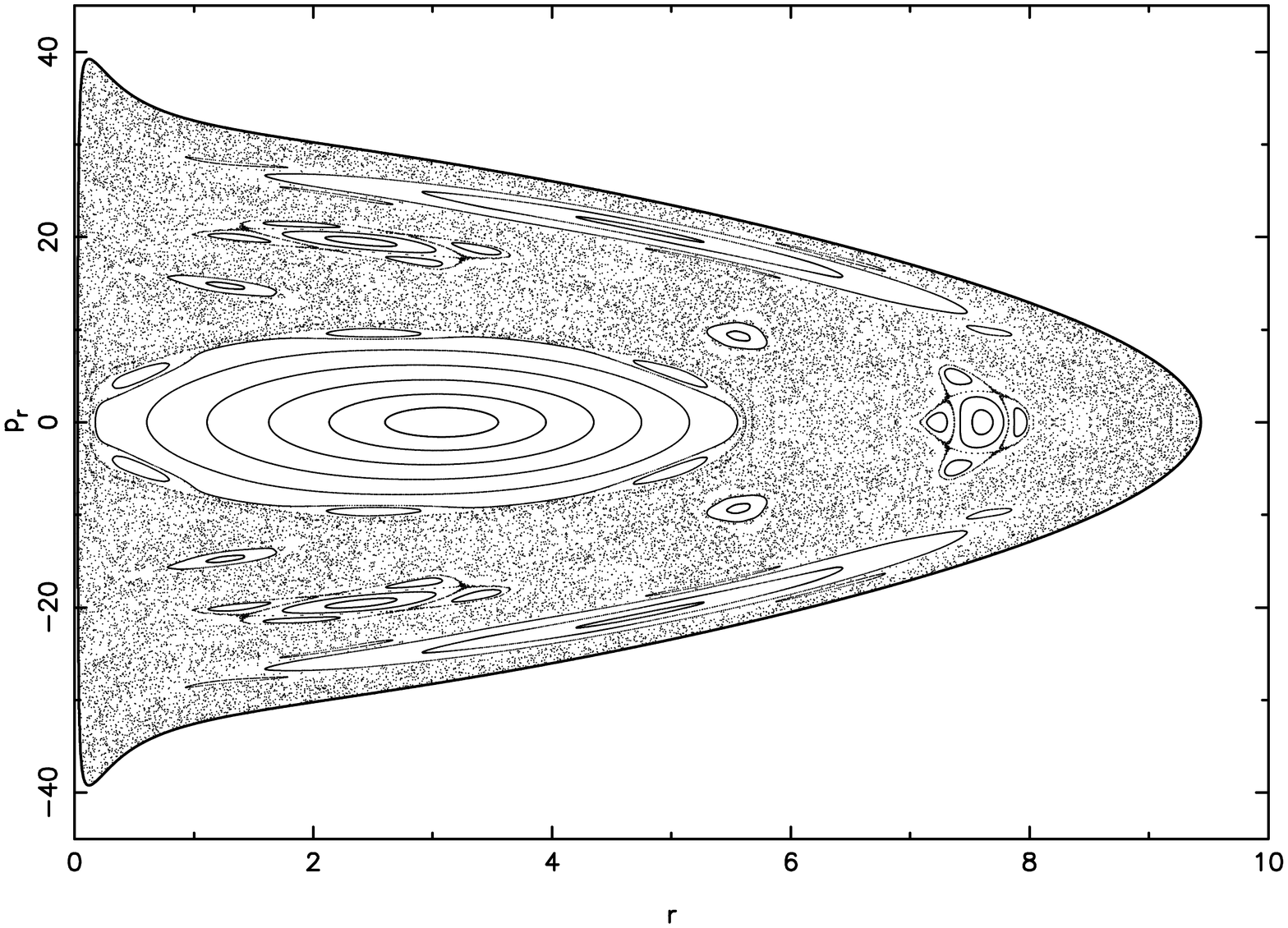}}\hspace{6cm}
                         \rotatebox{270}{\includegraphics*{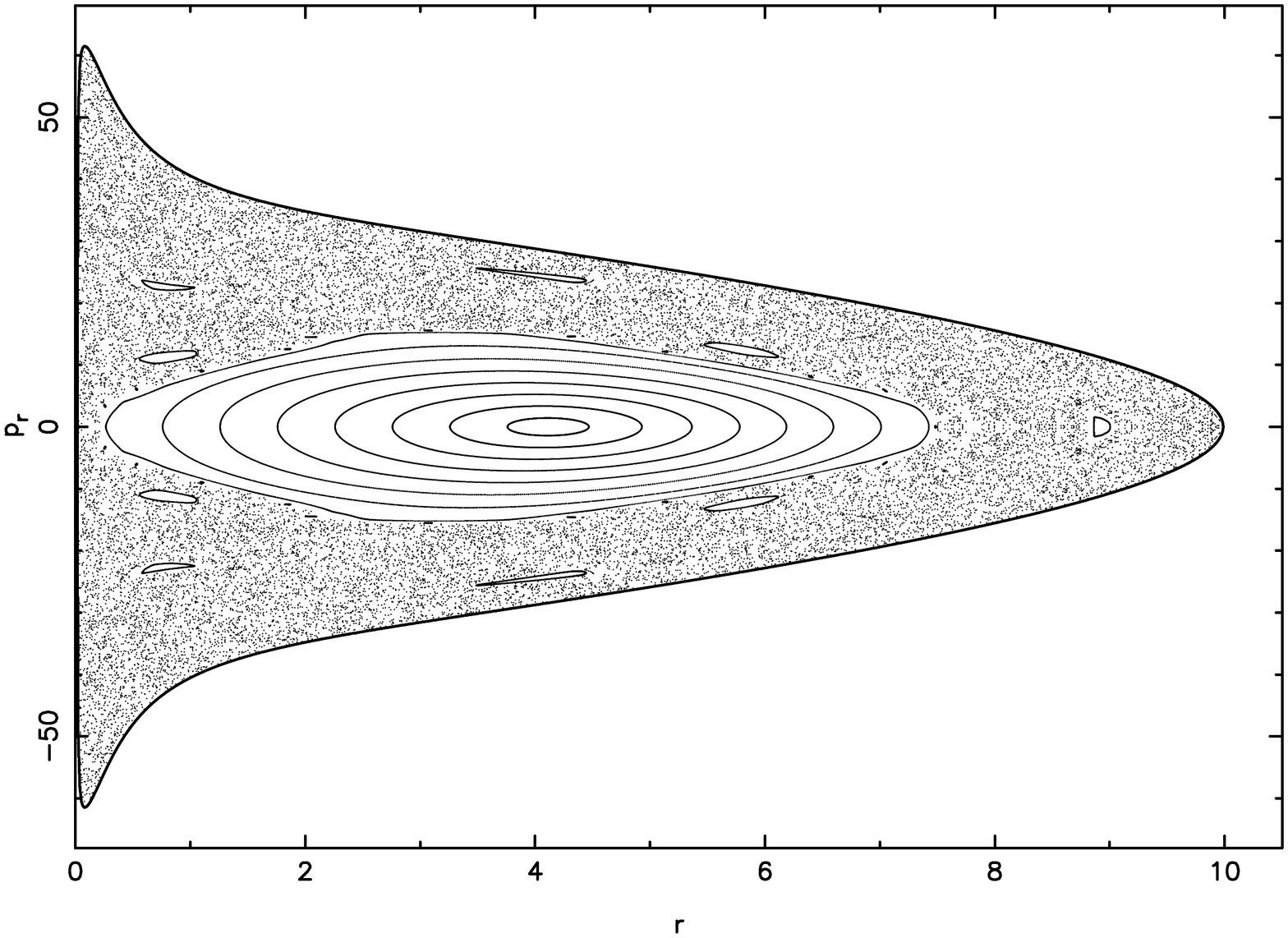}}}
\vskip 0.1cm
\caption{(a-b): The $(r,p_r)$ phase planes when $E=-950$ and $L_z=1$, while (a-left): $M_n=100$ and (b-right): $M_n=400$.}
\end{figure*}
\begin{figure*}[!tH]
\centering
\resizebox{0.8\hsize}{!}{\rotatebox{270}{\includegraphics*{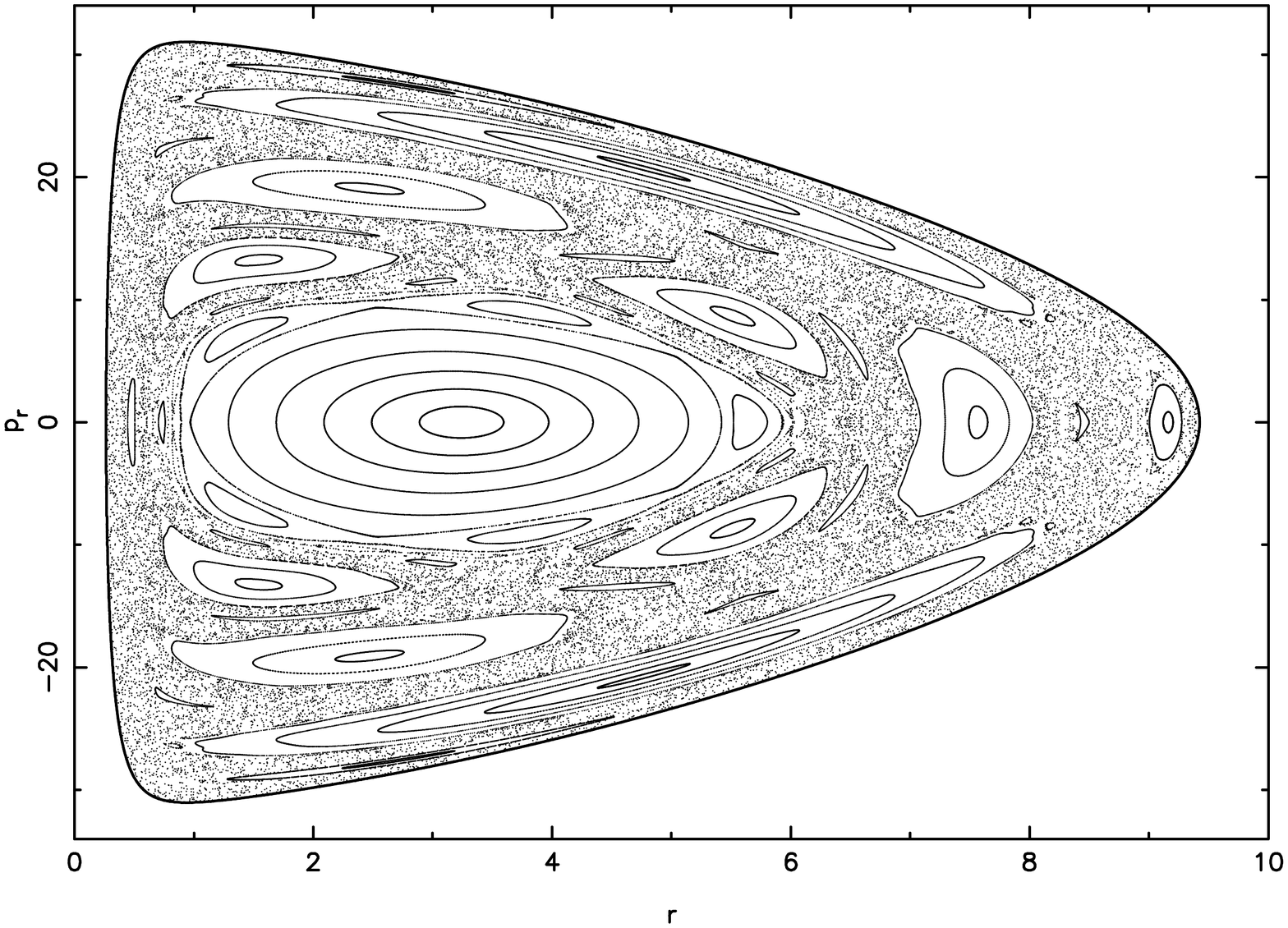}}\hspace{6cm}
                         \rotatebox{270}{\includegraphics*{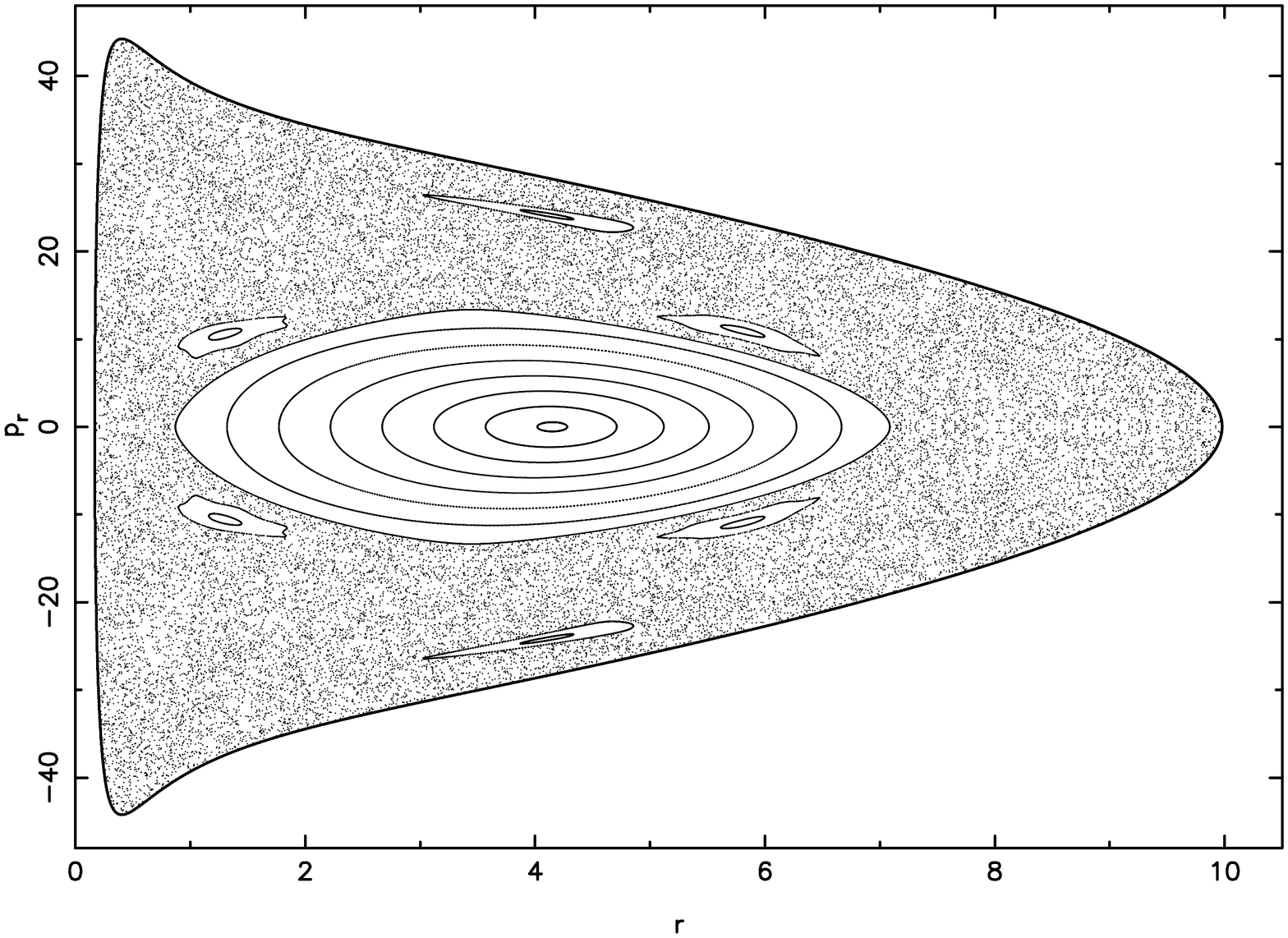}}}
\vskip 0.1cm
\caption{(a-b): The $(r,p_r)$ phase planes when $E=-950$ and $L_z=10$, while (a-left): $M_n=100$ and (b-right): $M_n=400$.}
\end{figure*}
\begin{figure*}[!tH]
\centering
\resizebox{0.8\hsize}{!}{\rotatebox{270}{\includegraphics*{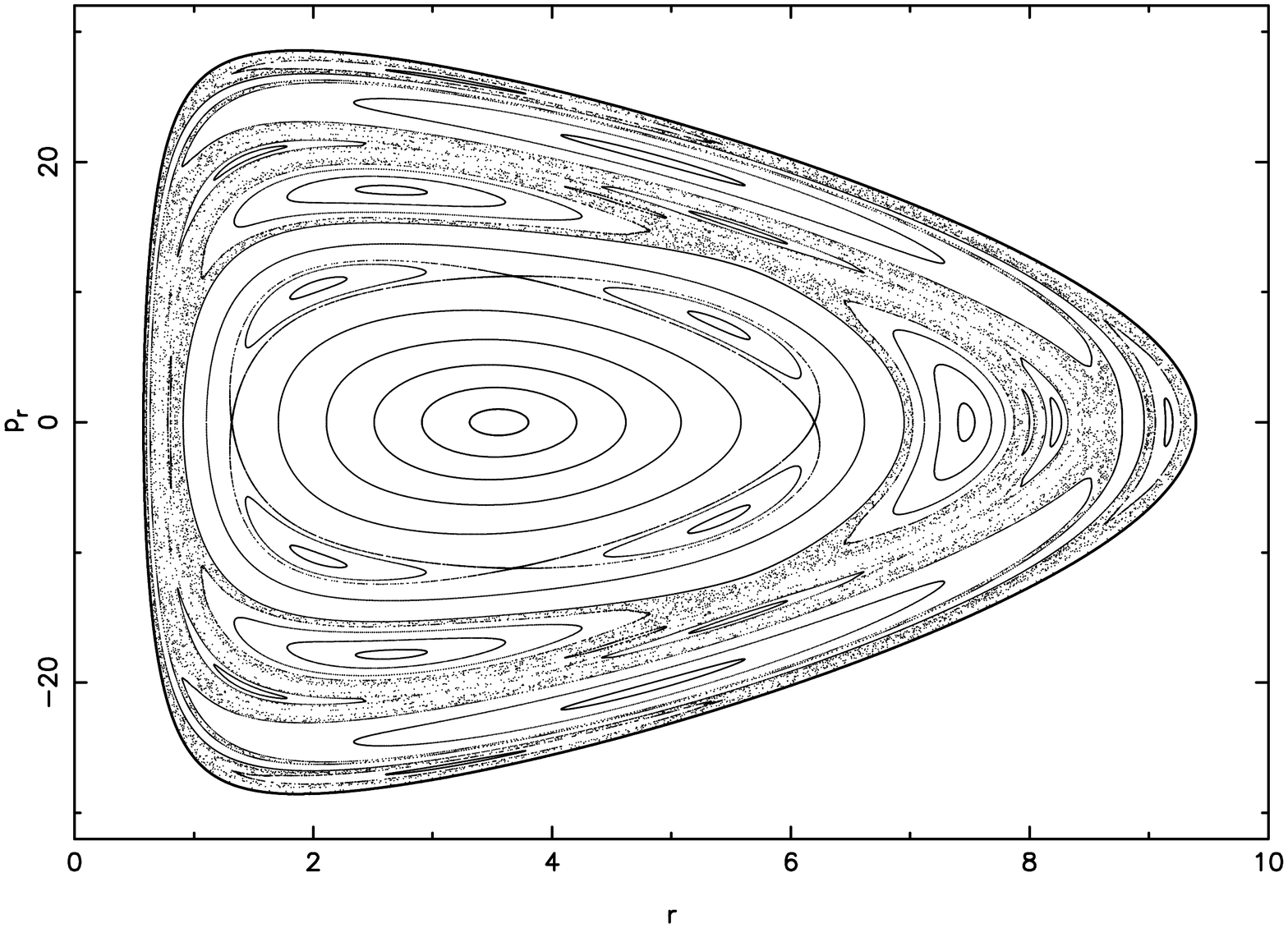}}\hspace{6cm}
                         \rotatebox{270}{\includegraphics*{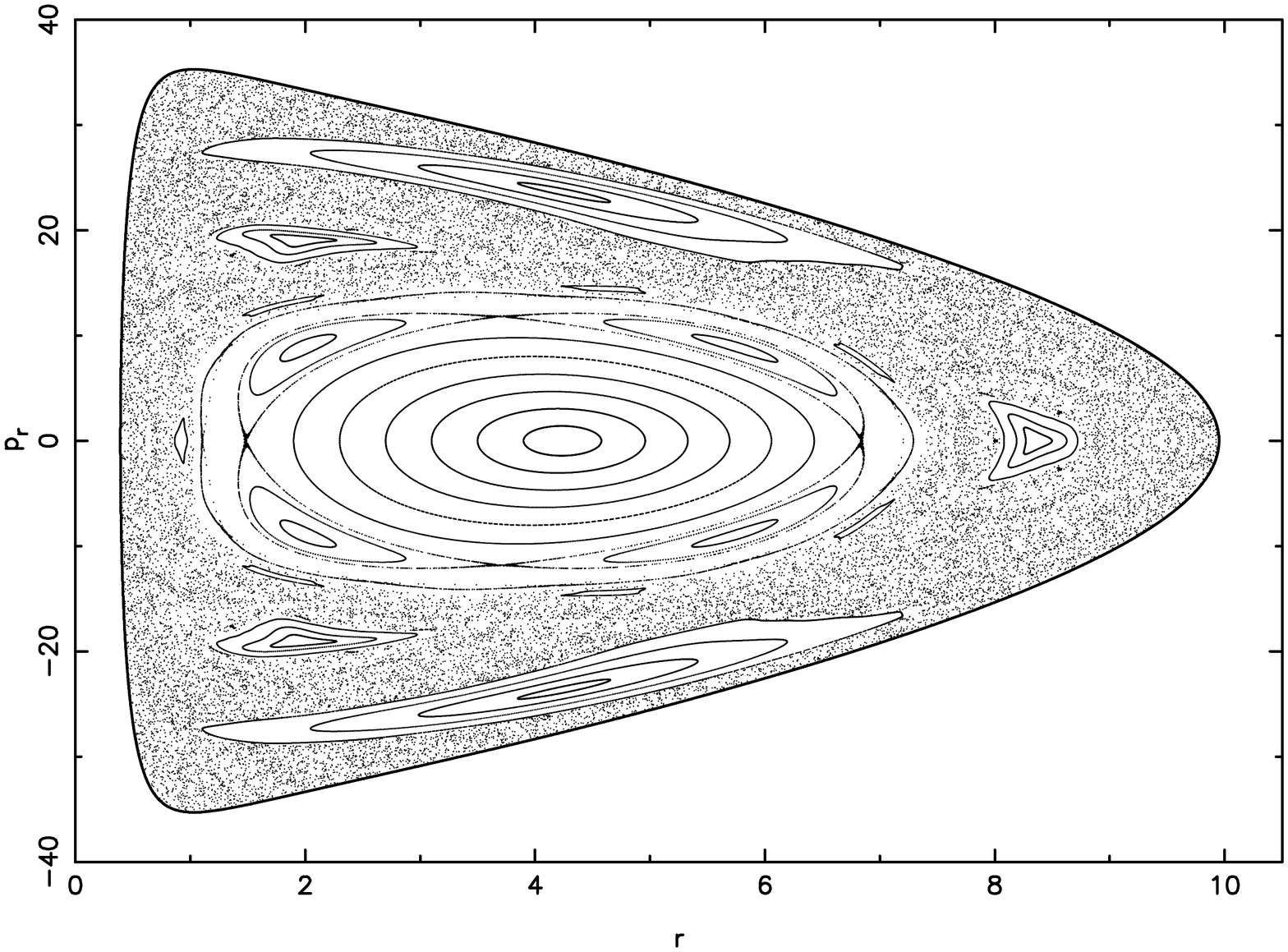}}}
\vskip 0.1cm
\caption{(a-b): The $(r,p_r)$ phase planes when $E=-950$ and $L_z=20$, while (a-left): $M_n=100$ and (b-right): $M_n=400$.}
\end{figure*}
\begin{figure*}[!tH]
\centering
\resizebox{0.8\hsize}{!}{\rotatebox{270}{\includegraphics*{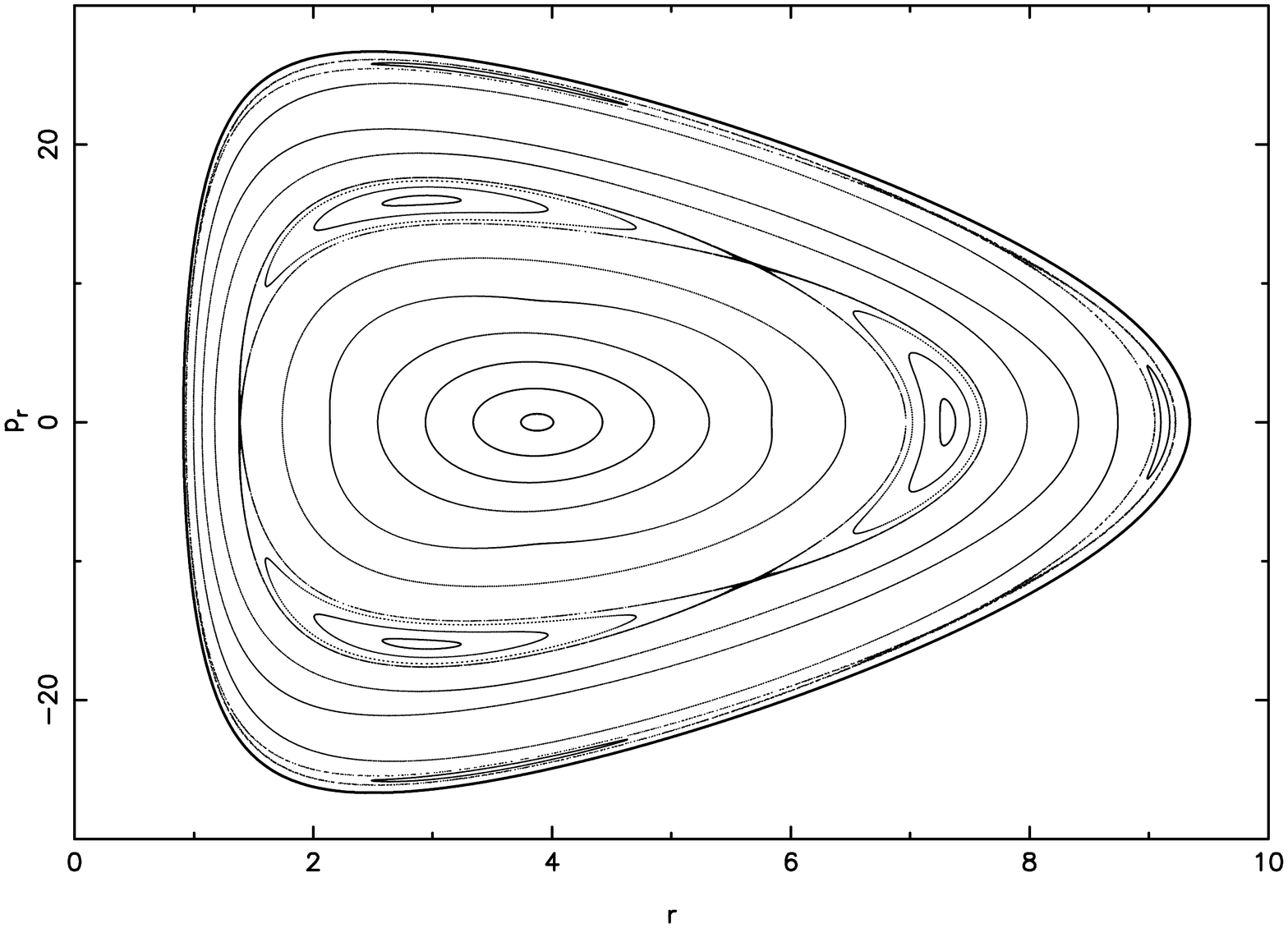}}\hspace{6cm}
                         \rotatebox{270}{\includegraphics*{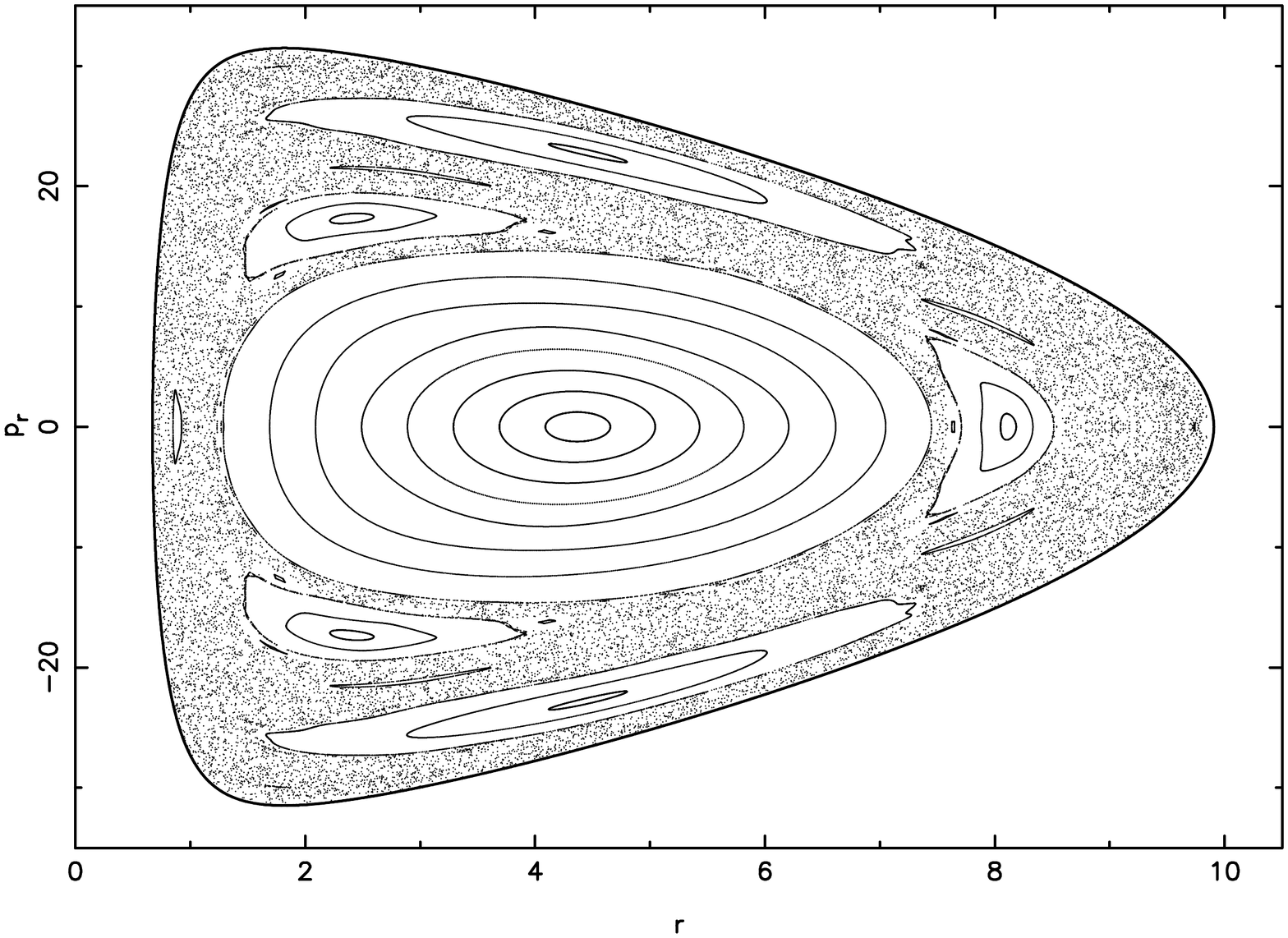}}}
\vskip 0.1cm
\caption{(a-b): The $(r,p_r)$ phase planes when $E=-950$ and $L_z=30$, while (a-left): $M_n=100$ and (b-right): $M_n=400$.}
\end{figure*}

In order to study the regular or chaotic nature of the orbits, we shall use the classical method of the $(r,p_r)$, $z=0, p_z>0$ Poincar\'{e} surface of section technique, where $p_r$ and $p_z$ are the radial and the vertical momenta respectively. If we set $z=p_z=0$ in equation (8), we obtain the limiting curve in the $(r,p_r)$ phase plane, which is the curve containing all the invariant curves, for a given value of the energy integral $E$. The limiting curve of the dynamical system corresponds to
\begin{equation}
\frac{1}{2}p_r^2 + \Phi_{eff}(r) = E \ \ \ .
\end{equation}
\begin{figure*}[!tH]
\centering
\resizebox{0.8\hsize}{!}{\rotatebox{0}{\includegraphics*{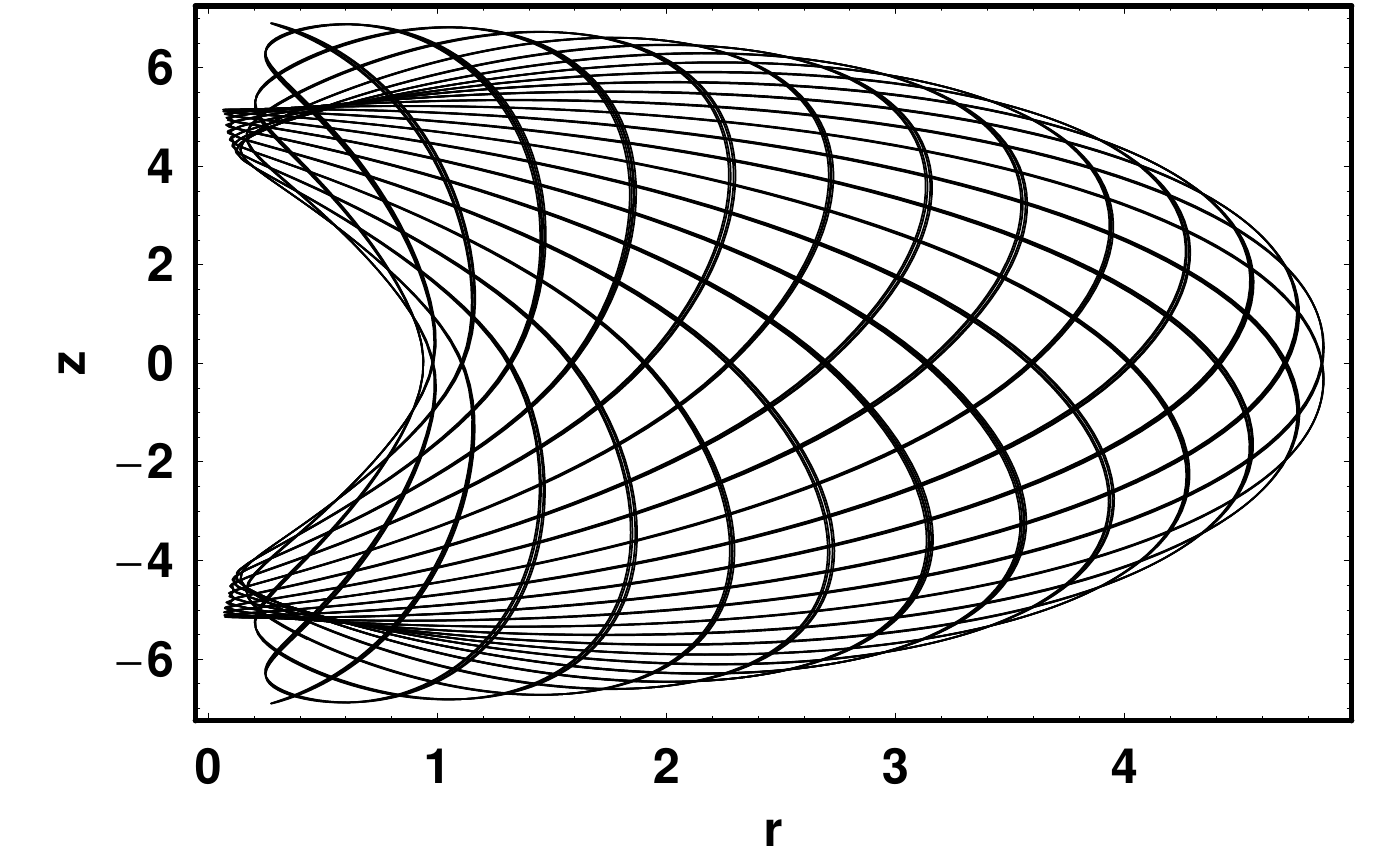}}\hspace{4cm}
                         \rotatebox{0}{\includegraphics*{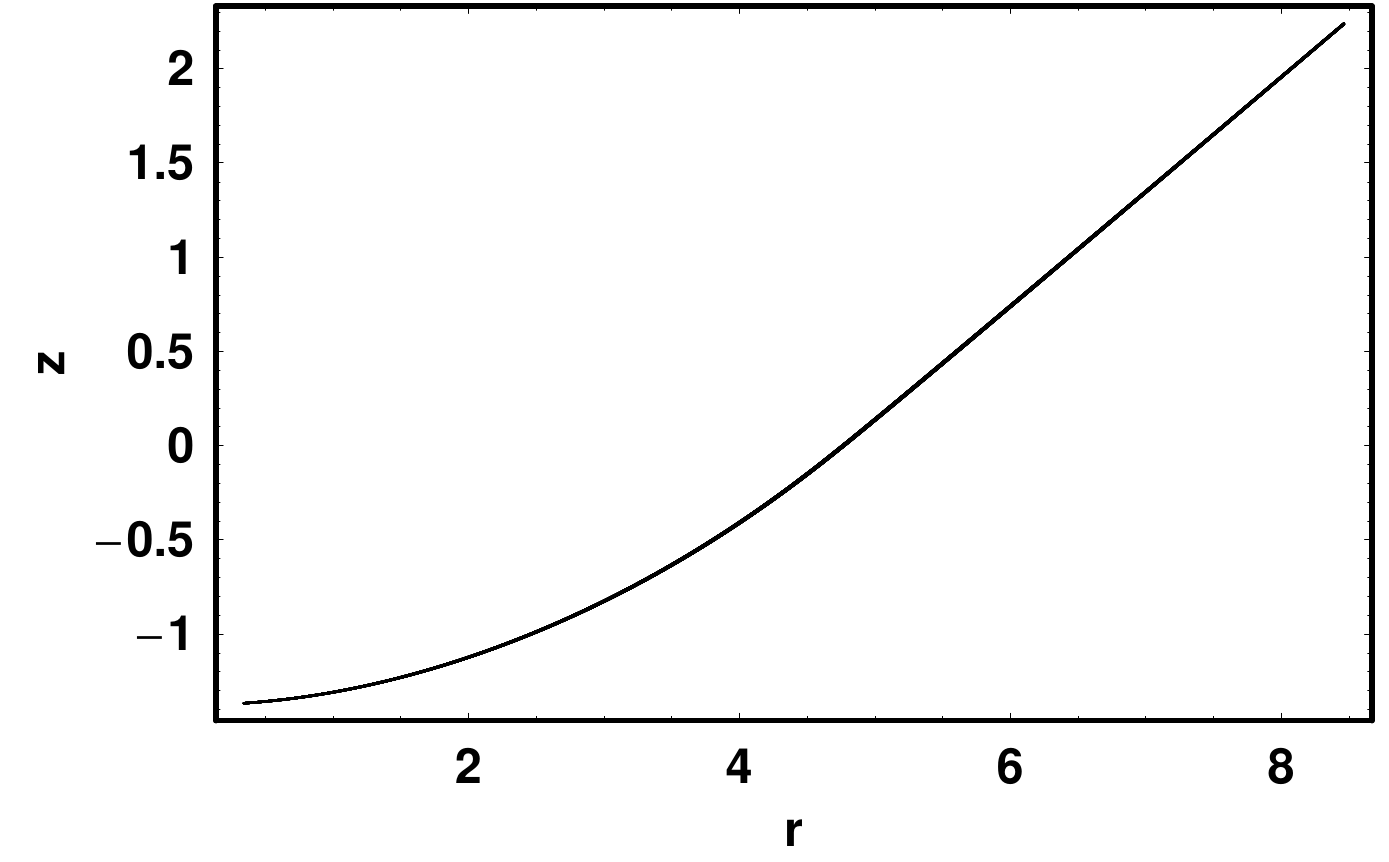}}}
\resizebox{0.8\hsize}{!}{\rotatebox{0}{\includegraphics*{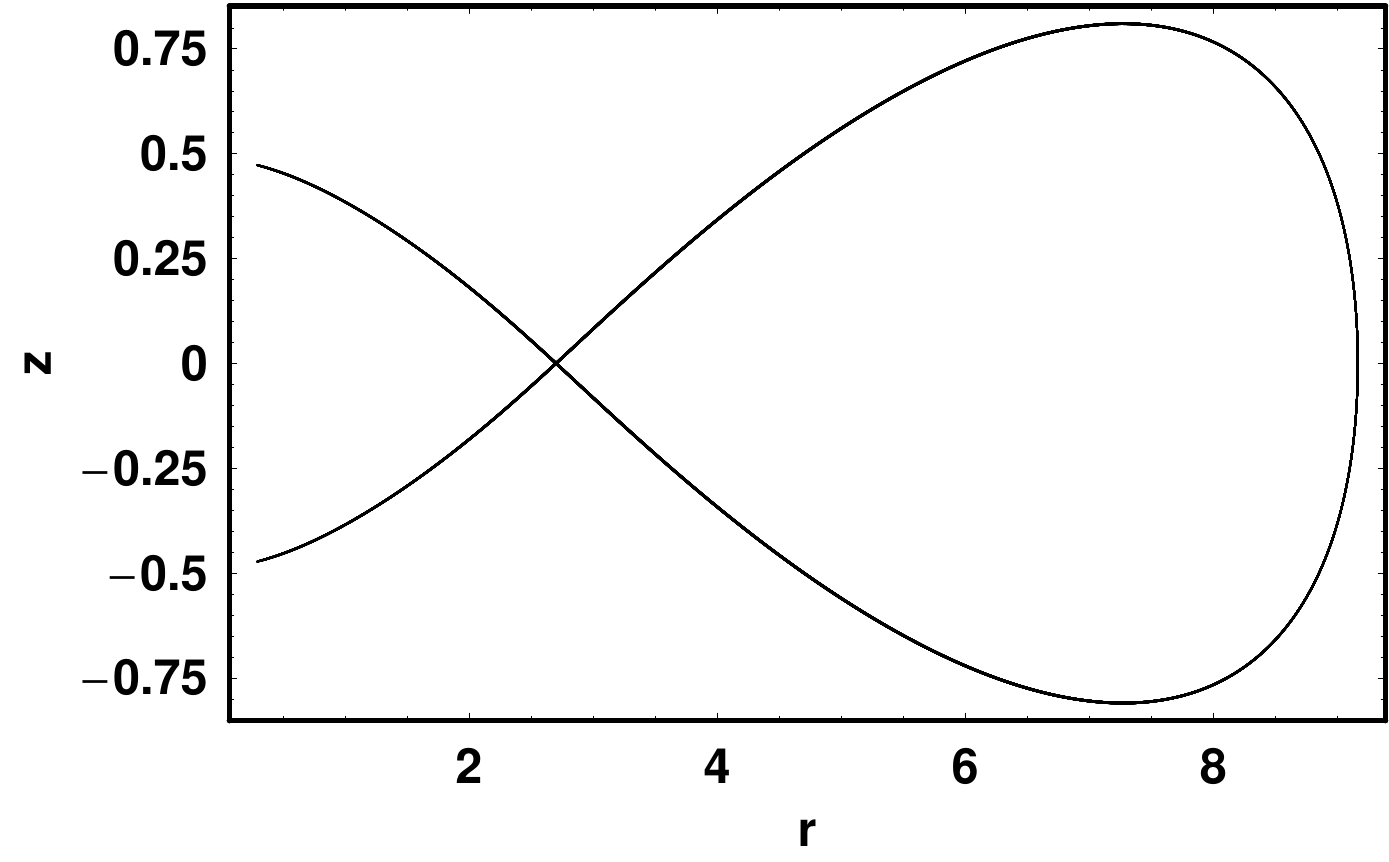}}\hspace{4cm}
                         \rotatebox{0}{\includegraphics*{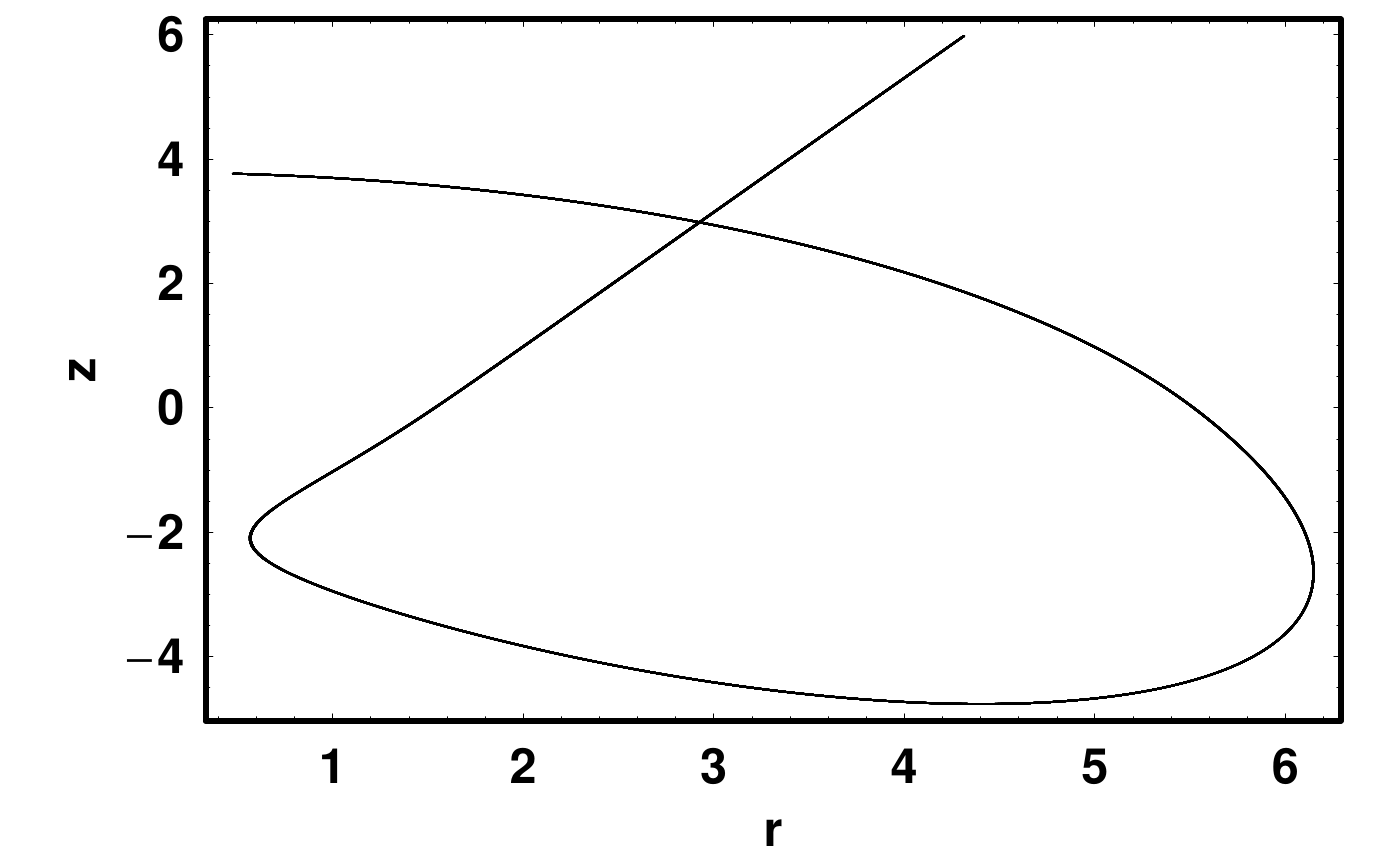}}}
\resizebox{0.8\hsize}{!}{\rotatebox{0}{\includegraphics*{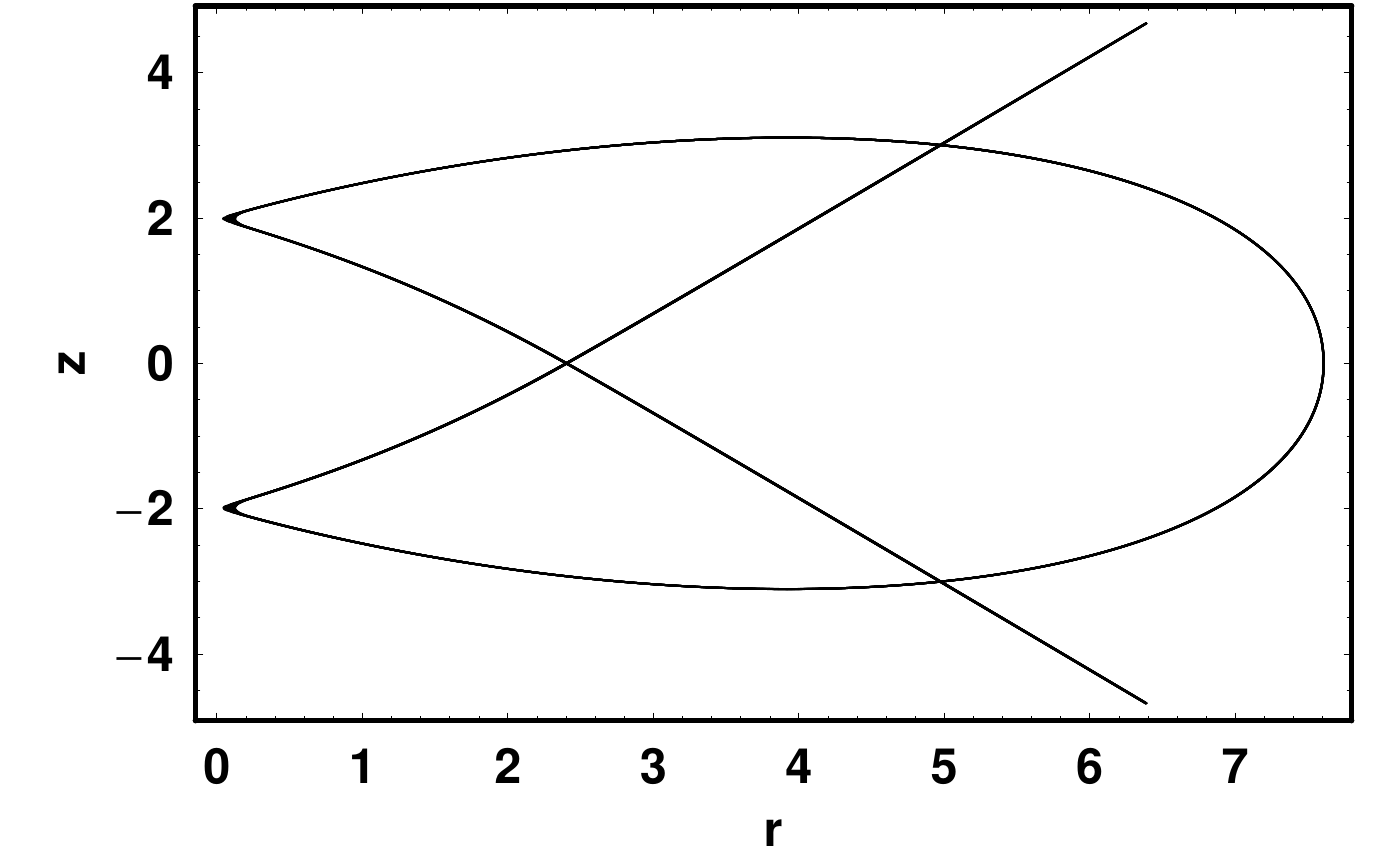}}\hspace{4cm}
                         \rotatebox{0}{\includegraphics*{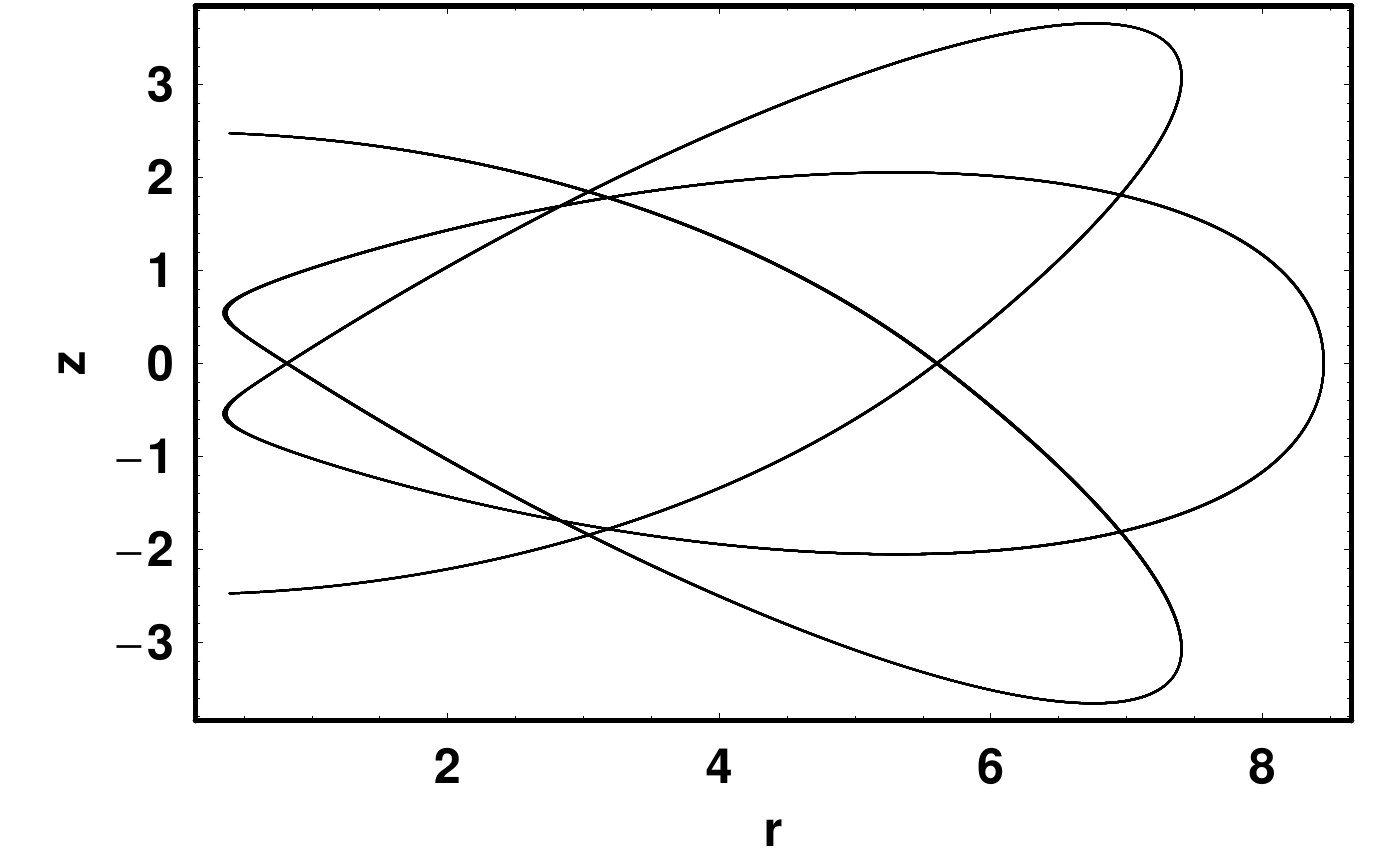}}}
\resizebox{0.8\hsize}{!}{\rotatebox{0}{\includegraphics*{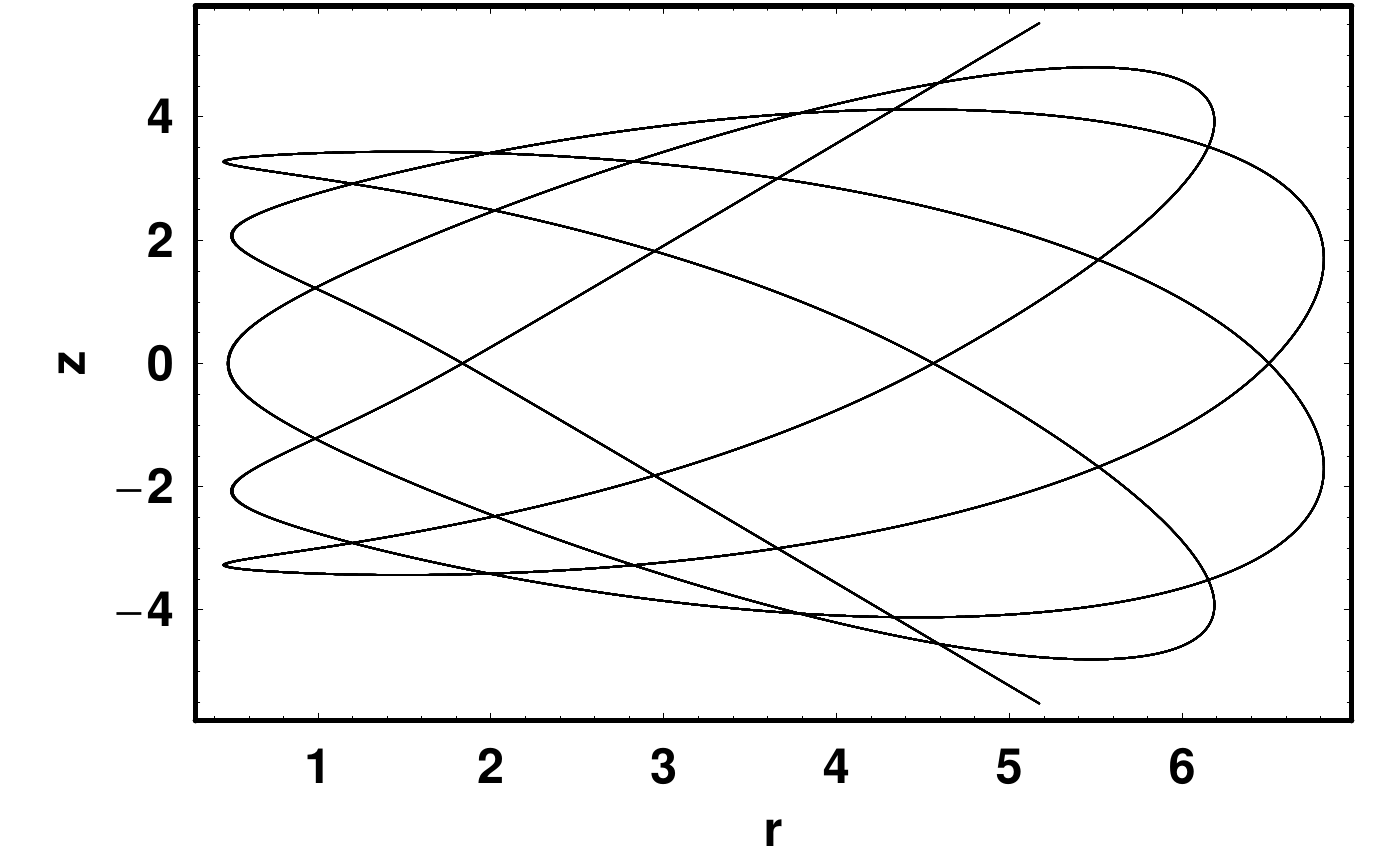}}\hspace{4cm}
                         \rotatebox{0}{\includegraphics*{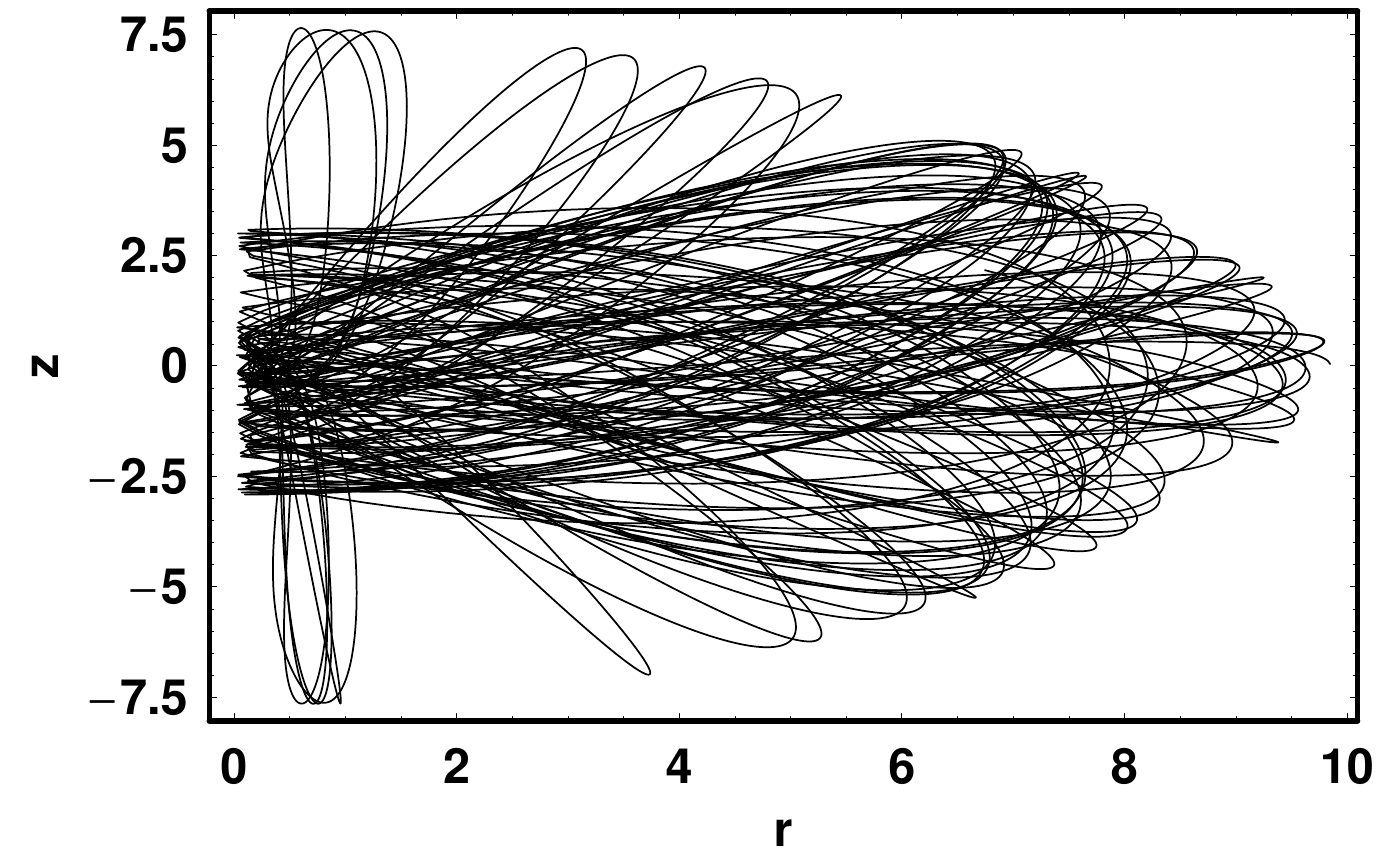}}}
\vskip 0.1cm
\caption{(a-h): Eight representative orbits of the dynamical system. The initial conditions and the values of all the parameters are given in the text.}
\end{figure*}

Figure 4a shows the $(r,p_r)$ phase plane when $L_z=1, M_n=100$ and $c_n=0.25$. As we see the majority of the phase plane is covered by chaotic orbits. Regular orbits are confined mainly near the central part of the phase plane. There are also several smaller islands embedded in the chaotic sea, which produced by secondary resonances. Figure 4b shows the $(r,p_r)$ phase plane when $L_z=1, M_n=400$ and $c_n=0.25$. In this case, where the nucleus is more massive, the extent of the chaotic region is larger, while the resonant phenomena are confined. Moreover, we observe that we have an increase in the radial velocity of the stars near the center of the galaxy. In Figure 5a we see the $(r,p_r)$ phase plane when $L_z=10, M_n=100$ and $c_n=0.25$. Here things are very different, since the majority of phase plane is covered by regular orbits now. Nevertheless a considerable chaotic area is present but it is confined mainly in the outer parts of the phase plane. Furthermore, we observe that the entire phase plane contains many sets of islands, which correspond to resonant orbits of higher multiplicity. In Figure 5b one can see the $(r,p_r)$ phase plane when $L_z=10, M_n=400$ and $c_n=0.25$. The main difference from Fig. 5a is that now the nucleus is more massive and the small chaotic area of Fig. 5a has become a vast chaotic sea. Furthermore, many of the sets of islands of invariant curves shown in Fig. 5a disappeared. Figure 6a shows the $(r,p_r)$ phase plane when $L_z=20, M_n=100$ and $c_n=0.25$. We observe that the extent of the chaotic orbits has been decreased drastically as the value of the angular momentum increases. On the contrary, from the $(r,p_r)$ phase plane shown in Figure 6b when $L_z=20, M_n=400$ and $c_n=0.25$, it is evident that the increase in mass of the nucleus leads to an increase of the chaotic region. We shall come to that later in this section. In Figure 7a we see the $(r,p_r)$ phase plane when $L_z=30, M_n=100$ and $c_n=0.25$. We observe that the entire phase plane is covered only by regular orbits. This is logical because the value $L_z=30$ of the angular momentum is larger than the critical value $L_{zc}=28$ (see Fig. 3). Figure 7b shows the $(r,p_r)$ phase plane when $L_z=30, M_n=400$ and $c_n=0.25$. As expected the increase of the mass of the nucleus has introduced chaotic areas in the $(r,p_r)$ phase plane. In all $(r,p_r)$ phase planes the value of the energy integral is the same and equals to $E=-950$.
\begin{figure}[!tH]
\centering
\resizebox{0.95\hsize}{!}{\rotatebox{0}{\includegraphics*{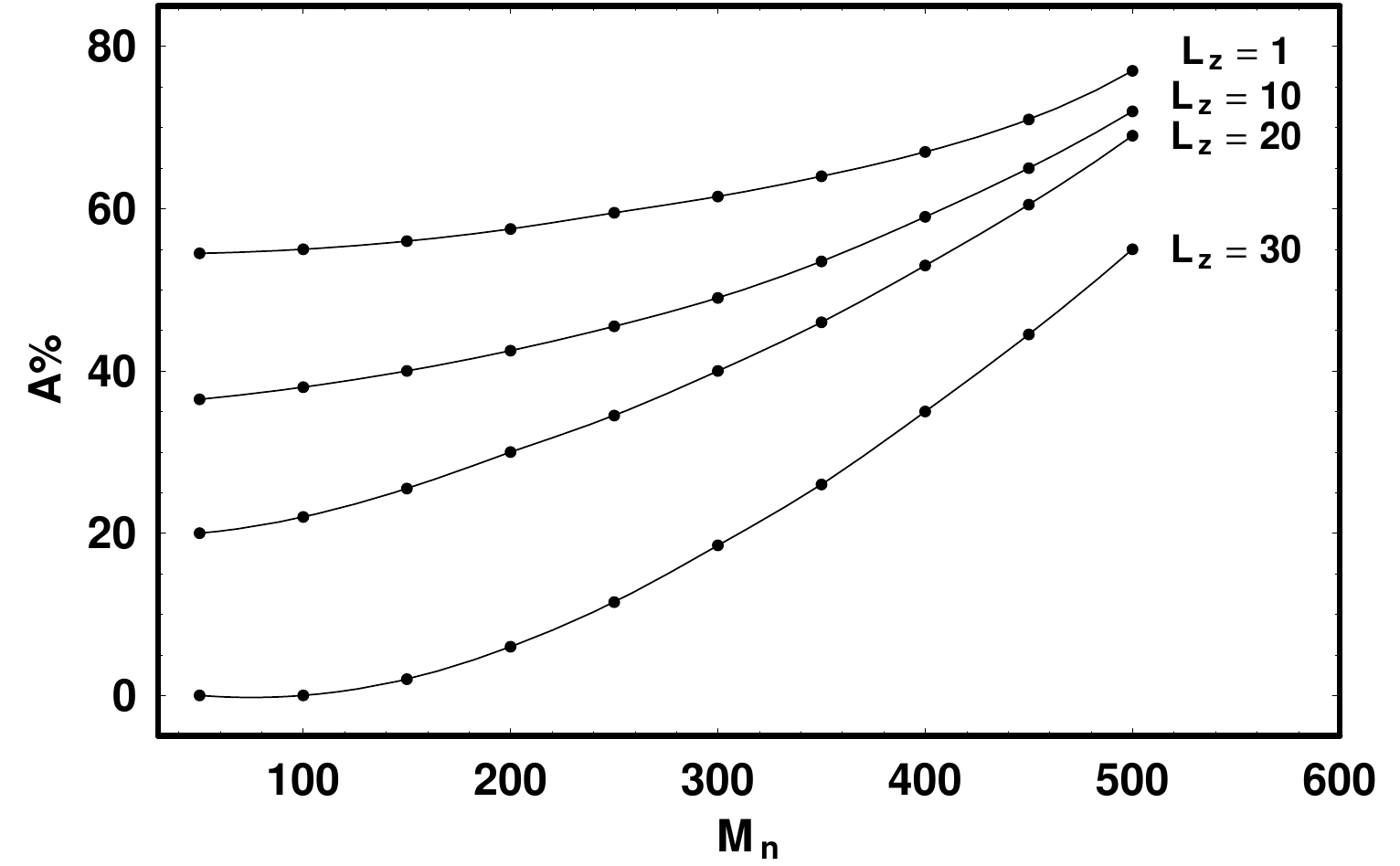}}}
\caption{A plot of the relation between the percentage of the area $A\%$ in the $(r,p_r)$ phase planes covered by chaotic orbits and the mass of the nucleus $M_n$.}
\end{figure}
\begin{figure}[!tH]
\centering
\resizebox{0.95\hsize}{!}{\rotatebox{0}{\includegraphics*{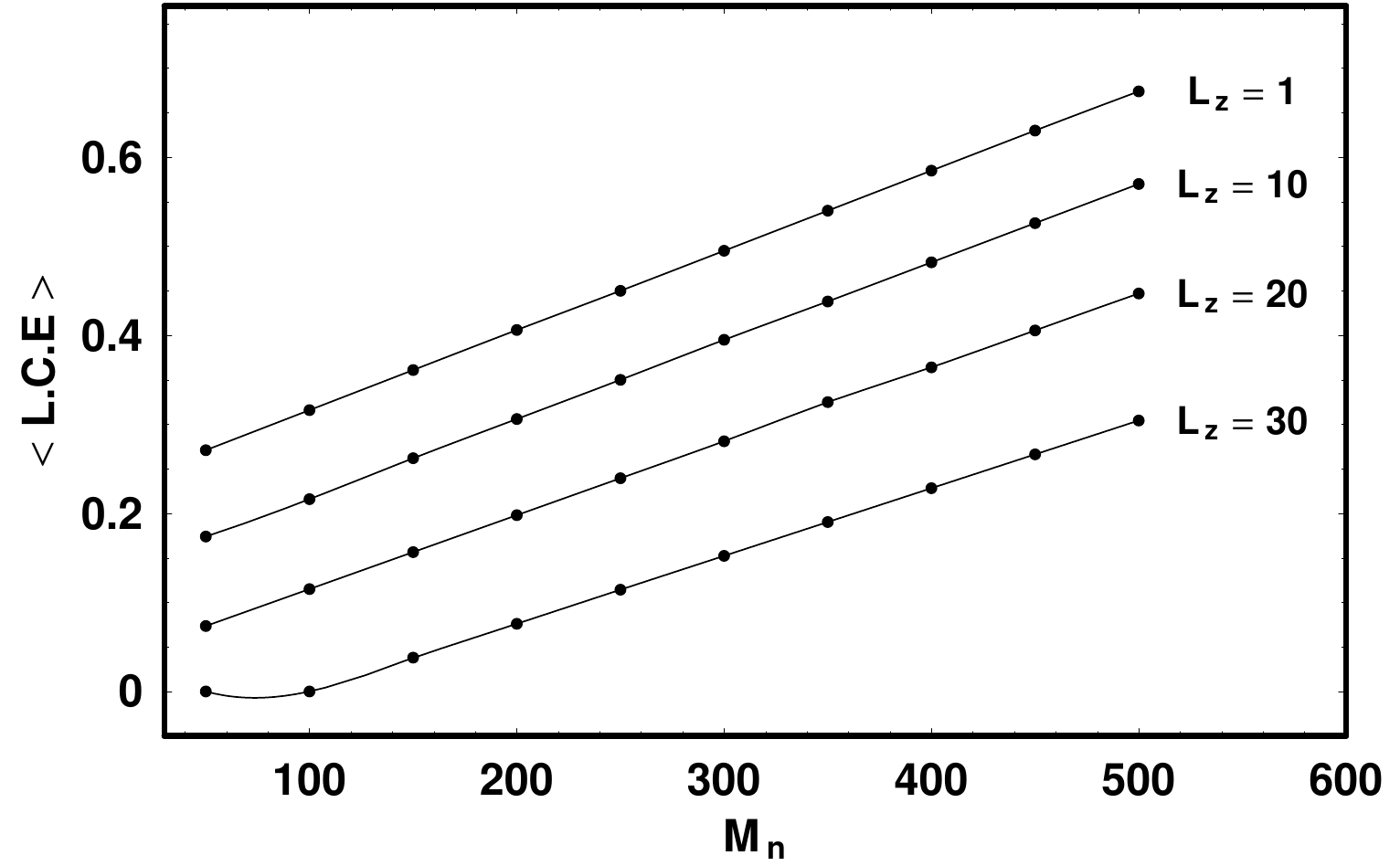}}}
\caption{A plot of the relation between the average value of the L.C.E and the mass of the nucleus $M_n$.}
\end{figure}

We observe from Figs. 4-7, that the extent of the chaotic regions in the phase plane increases as the nucleus becomes more massive. On the other hand, as we increase the value of the angular momentum $L_z$, the chaotic regions become smaller. Therefore, we conclude that the portion of chaotic orbits with low or high values of angular momentum, in disk galaxies is larger when a massive and dense nucleus is present. Here we must notice that the whole area of the $(r,p_r)$ phase plane is reduced and becomes smaller as $L_z$ increases. The outermost solid curve in the $(r,p_r)$ phase planes shown in Figs. 4-7 is the limiting curve.

Figure 8a-h shows eight representative orbits of the dynamical system. Figure 8a shows a regular orbit of 1:1 resonance when $L_z=1, M_n=100, c_n=0.25$. The initial conditions are: $r_0=0.94, z_0=0, p_{r0}=0$. The value of $p_{z0}$ is found always from the energy integral (8) for all orbits, while the energy integral has a constant value, which is $E=-950$. In Figure 8b we see a periodic orbit, which is characteristic of the 2:1 resonance. For this orbit we have $L_z=10, M_n=100, c_n=0.25$, while the initial conditions are: $r_0=4.78, z_0=0, p_{r0}=20.7$. This orbit carries stars out of the galactic plane. Figure 8c shows a periodic orbit of the 2:3 resonance, when $L_z=10, M_n=100, c_n=0.25$. The initial conditions are: $r_0=9.164, z_0=0, p_{r0}=0$. In Figure 8d one observes a periodic orbit when $L_z=10, M_n=100, c_n=0.25$, with initial conditions: $r_0=1.54, z_0=0, p_{r0}=13.25$. Figure 8e shows a periodic orbit which is characteristic of the 4:3 resonance. Here $L_z=1, M_n=100, c_n=0,25$, while the initial conditions are: $r_0=7.61, z_0=0, p_{r0}=0$. In Figure 8f a periodic orbit of the 3:5 resonance family is shown. In this case we have $L_z=10, M_n=100, c_n=0.25$ and the initial conditions are: $r_0=8.455, z_0=0, p_{r0}=0$. Figure 8g depicts a complicated periodic orbit of higher multiplicity. Here $L_z=10, M_n=100, c_n=0.25$, while the initial conditions are: $r_0=0.476, z_0=0, p_{r0}=0$. A chaotic orbit is shown in Figure 8h, when $L_z=1, M_n=400, c_n=0.25$ and the initial conditions are: $r_0=9.85, z_0=0, p_{r0}=0$. Note that stars moving in chaotic orbit as this shown in Fig. 8h, are scattered to the halo when approaching the massive nucleus. All orbits were calculated for a time period of 100 time units. We can observe that the periodic orbits shown in Figs. 8c, 8e, 8f and 8g stay relatively close to the galactic plane and therefore, the corresponding families of regular orbits, support the disk structure of the galaxy.

Figure 9 shows the percentage of the area $A\%$ covered by chaotic orbits in the $(r,p_r)$ phase planes as a function of the mass of the nucleus $M_n$, for four different values of the angular momentum $L_z$. We observe that the percentage $A\%$ increases as the mass of the nucleus $M_n$ increases in four cases. We must point out that from this diagram we can extract very useful information. In particular, low angular momentum stars, display strong chaotic behavior. Moreover, as we increase the value of the angular momentum, the increase of the chaotic percentage, as the nucleus gaining mass becoming more massive, is more intense. It seems that when the nucleus is massive enough, $M_n \geq 500$, the value of the angular momentum cannot affect the nature of the orbits due to the fact that $M_n$ is the dominant parameter now. We see in Fig. 9 that all four fitting curves, which correspond to the four different values of the $L_z$, converge when $M_n \geq 500$. We must point out, that the percentage $A\%$ is calculated as follows: we choose 1000 orbits with different and random initial conditions $(r_0,p_{r0})$ in each phase plane and then divide the number of those who produce chaotic orbits to the total number of orbits.

In Figure 10 we see a plot of the average value of the Lyapunov Characteristic Exponent - L.C.E (see Lichtenberg and Lieberman, 1992) as a function of the mass of the nucleus $M_n$, for four different values of the angular momentum $L_z$. We can observe that the $< L.C.E >$ increases linearly, as the mass of he nucleus $M_n$ increases in four cases. Furthermore, we see that the fitting curves are almost parallel lines. We conclude, that the degree of chaos increases linearly as the mass of the nucleus increases. In addition, low angular momentum stars display larger degree of chaoticity than high angular momentum stars for the same mass of the nucleus. We shall try to explain these results in the next section. Here we must notice, that it is well known that the value of L.C.E is different in each chaotic component (see Saito \& Ichimura, 1979). As we have in all cases regular regions and only one unified chaotic region in each $(r,p_r)$ phase plane, we calculate the average value of L.C.E by taking 500 orbits with different and random initial conditions $(r_0,p_{r0})$ in the chaotic region in each case. Note that, all calculated L.C.Es were different on the fifth decimal point in the same chaotic region.
\begin{figure}[!tH]
\centering
\resizebox{0.95\hsize}{!}{\rotatebox{0}{\includegraphics*{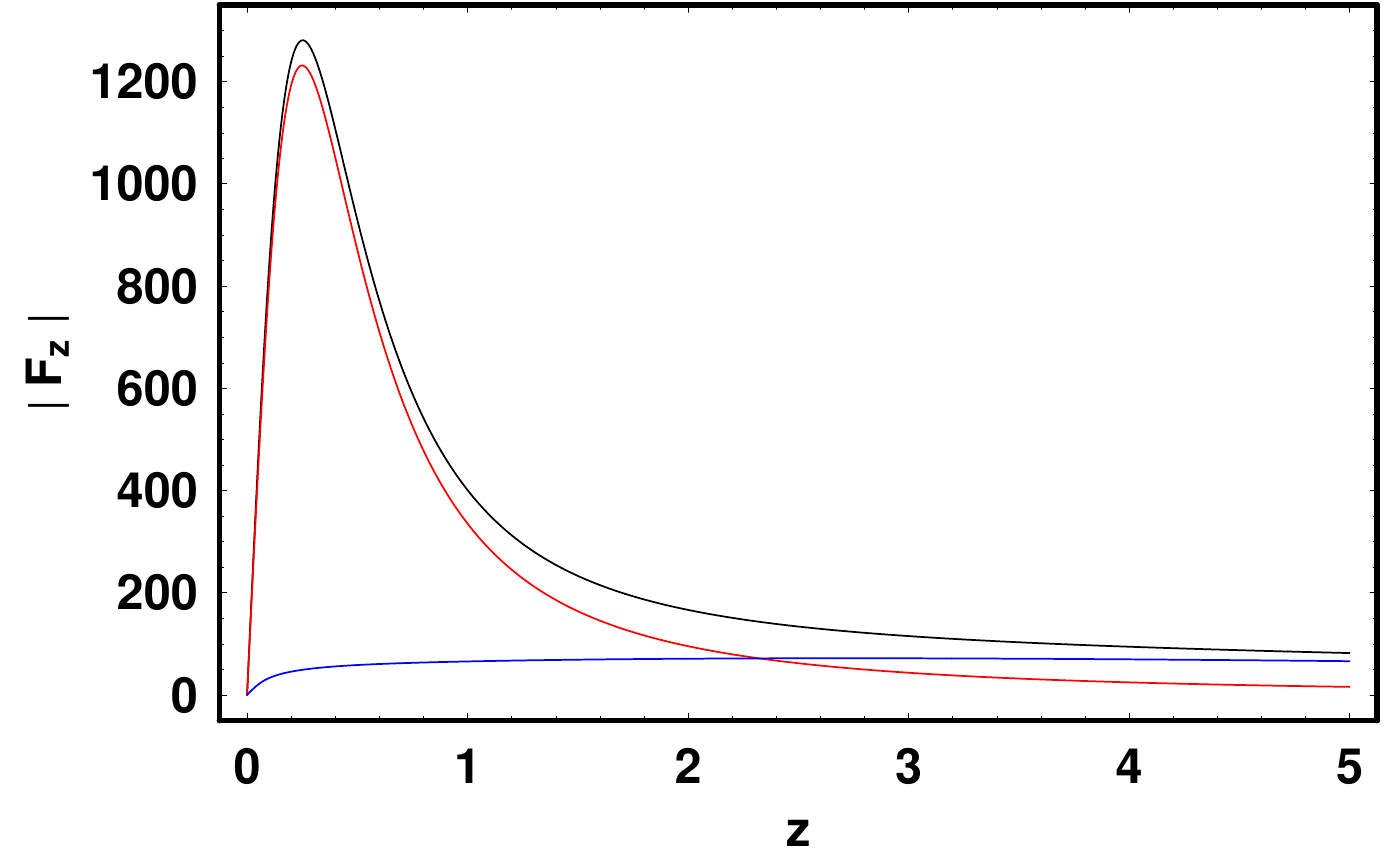}}}
\caption{A semi theoretical plot of the relation between $|F_z|$ and $M_n$. The black line represents the total $|F_{zt}|$ force, the red line is the contribution from the spherical nucleus $|F_{zn}|$, while the blue line is the contribution from the disk-halo potential, $|F_{zd}|$.}
\end{figure}
\begin{figure}[!tH]
\centering
\resizebox{0.95\hsize}{!}{\rotatebox{0}{\includegraphics*{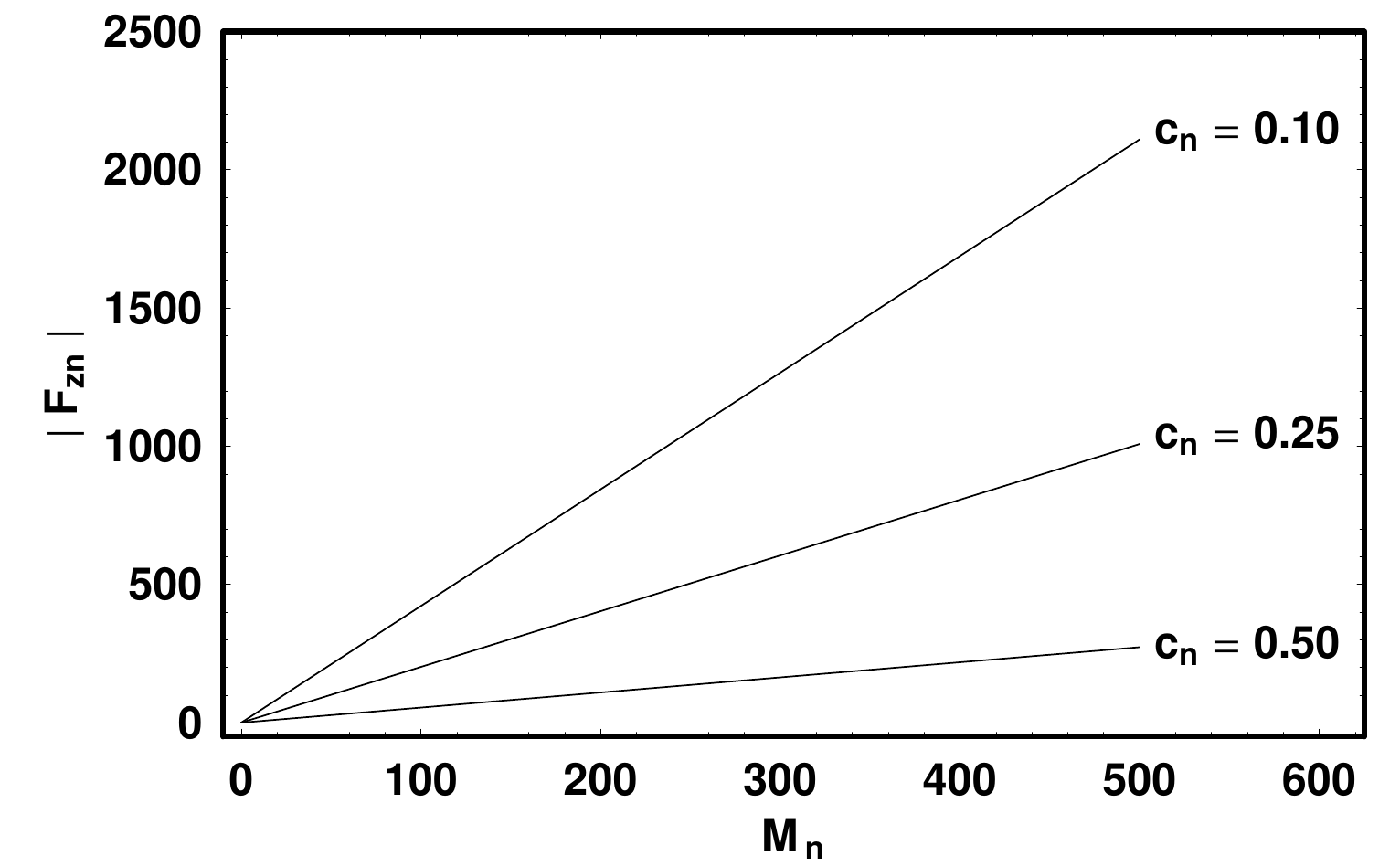}}}
\caption{A semi theoretical plot of the relation between the nuclear $|F_{zn}|$ force and the mass of the nucleus $M_n$, for three different values of the scale length $c_n$.}
\end{figure}

\section{A semi-theoretical approach}

In the present Section, we shall present some semi-theoretical arguments, together with elementary numerical calculations, in order to explain the numerically obtained relationships, given in Figs. 3, 9 and 10. In order to explain the results shown in Fig. 3, we use a mathematical analysis, similar to that used in Zotos (2011a,b). Our numerical simulations indicate that, near the nucleus, the $F_z$ force component is dominant (it is approximately 20 times grater than at $r=8.5$ kpc, for the same value of $z$) and in fact in nearly equal to that produced by the nucleus itself. The total $|F_{zt}|$ force when $M_n=400$ and $c_n=0.25$, is shown as the black line in Figure 11. In the same plot, the red line is the contribution from the spherical nucleus, $|F_{zn}|$, while the blue line is the contribution from the disk-halo component, $|F_{zd}|$. Even when $z$ is of order of unity, about $82\%$ of the total $|F_{zt}|$ force is due to the massive nucleus. Only when $z \geq 2.4$ kpc the $|F_{zd}|$ component becomes the dominant vertical force. Furthermore, it is the nuclear force that causes the scattering to the halo leading to chaotic motion. In fact, on approaching the nucleus. there is a change in the star's momentum in the $z$ direction, given by
\begin{equation}
m \Delta \upsilon_z = < F_z > \Delta t \ \ \ ,
\end{equation}
where $m$ is the mass of the star, $< F_z >$ is the total (i.e., nuclear) average $z$ force, while $\Delta t$ is the duration of the encounter. It was observed, that the star's deflection into the halo, proceeds in each case cumulatively, a little more with each successive pass by the nucleus, rather than with a single ``dramatic" encounter. It is assumed, that the star is scattered off the galactic plane and goes to the halo, after $n$ $(n>1)$ encounters, when the total change in the momentum in the $z$ direction, is of order of $m \upsilon_\phi$, where $\upsilon _{\phi}=L_z/r$, is the tangential velocity of the star near the nuclear region. Therefore, we shall assume that
\begin{equation}
m\sum\limits_{i=1}^{n}{\Delta {{\upsilon }_{zi}}\approx \left\langle {{F}_{z}} \right\rangle }\sum\limits_{i=1}^{n}{\Delta {{t}_{i}}} \ \ \ .
\end{equation}
Setting
\begin{eqnarray}
m \sum\limits_{i=1}^{n} \Delta \upsilon _{zi}&=& m \upsilon_{\phi} = m \frac{L_z}{r}, \nonumber \\
\sum\limits_{i=1}^{n} \Delta t_i&=&T_e, \nonumber \\
\frac{< F_z >}{m} &=& F_{zn}, \nonumber \\
m&=&1,
\end{eqnarray}
in Eq. (13) we obtain
\begin{equation}
\frac{{{L}_z}}{r}\approx < F_{zn} > {{T}_{e}} \ \ \ .
\end{equation}

Initially, if $\upsilon_z \geq \upsilon _ {\phi}$ the orbit is chaotic. Otherwise, $\upsilon_z \ll \upsilon _ {\phi}$ and after successive encounters with the galactic center, $\upsilon_z$ will increase, until it reaches $\upsilon_{\phi}$. Then the orbit will be chaotic. The time needed to reach the condition $\upsilon_z > \upsilon_{\phi}$ is $T > T_e$, where $T_e$ is the sum of the encounter durations (14). Let $T_f$ be the age of the galaxy, which corresponds here to the final time of numerical integrations. If $T_e > T_f$, then some stars do not have enough time to be scattered. We conclude from Eq. (15), that if $L_z > L_{zc}$, where $L_{zc}$ is the angular momentum computed with $T_e = T_f$, then there is still exist non-chaotic motions. Whereas for $L_z < L_{zc}$, all stars may have time to deploy chaotic motions.

For a given value of $r_0$ (the particular value is irrelevant as soon as $r_0$ is of order of unity), the quantitative behavior of $|F_z|$ force is shown in Fig. 11. For small values of $z$, $F_z$ behaves as a harmonic oscillator force, while for larger values of $z$, falls as $M_n/z^{3/2}$. Expanding the potential of the nucleus around the point $P\left(r=r_0,z=0\right)$, the square of the frequency of the harmonic motion in the $z$ direction is $M_n/\left(r_0^2 + c_n^2\right)^{3/2}$. Substituting the harmonic motion law in Eq. (15) with $r=r_0=z=c_n$, we find
\begin{equation}
M_n \approx 2 c_n L_{zc} \sqrt{2}/T_{ea} \ \ \ ,
\end{equation}
where, for convenience, we have written $T_{ea}$ instead of $T_e$. If we set $< F_z > = M_n/z^{3/2}, r=c_n, z=k c_n$ $(k\geq 2)$ into Eq. (15), we obtain
\begin{equation}
M_n \approx L_{zc}\sqrt{c_n k^3}/T_e \ \ \ .
\end{equation}

Equations (16) and (17) are very simplistic, but nevertheless, reproduce the observed linear relationship between $L_{zc}$ and $M_n$, which is shown in Fig. 3. Relation (17) also explains the dependence of the slope on the scale length of nucleus $c_n$. The larger is the $c_n$ the higher is the slope. This is because $T_e$ is observed to be the about the same for the same scale length of the nucleus. On the other hand, when $L_{zc}$ is very small, the star goes much closer to the nucleus, so that relation (16) must be used. Note that for the nucleus we always have $c_n < 1$ and because of the smaller mass of the nucleus $T_{ea} > T_e$. Therefore, a lower slope results for the line in the region of very low angular momenta or, what is equivalent, for very low nuclear masses (see Fig. 3).

The $F_{zn}$ nuclear force is
\begin{equation}
F_{zn} = - \frac{\partial \Phi_n(r,z)}{\partial z} = - \frac{M_n z}{\left(r^2 + z^2 + c_n^2\right)^{3/2}} \ \ \ .
\end{equation}
Figure 12 shows a plot of the nuclear $|F_{zn}|$ force as a function of the mass of the nucleus $M_n$, for three different values of the scale length $c_n$. As the scattering of the stars occurs near the nucleus, it must be at $r<1$ and $z<1$. The particular values of $r$ and $z$ are irrelevant. In this case, we choose: $r=r_0=c_n=0.25$ kpc and $z=z_0=0.10$ kpc. We observe that $|F_{zn}|$ increases linearly with $M_n$, in all cases. One can see, that the pattern between $F_{zn}$ and the mass of the nucleus (shown in Fig. 12), is similar to those between the chaotic percentage $A\%$ or the average value of the L.C.E and the mass of the nucleus $M_n$ (shown in Figs. 9 and 10 respectively), which have been obtained numerically. In both cases numerical outcomes suggest that, as the value of the mass of the nucleus $M_n$ increases, the percentage and also the degree of chaos in the dynamical system increases as well. Numerically found relationships given in Figs. 9 and 10 can be explained because, as we have said earlier, the $F_{zn}$ force is the main contributor, to the chaotic motion observed in the $(r,p_r)$ phase planes of the dynamical system. Moreover, from Fig. 12 one can conclude that when the scale length of the nucleus $c_n$ is smaller, that is when the nucleus is more dense, the $F_{zn}$ force is stronger, for each particular value of the mass of the nucleus $M_n$.

A large number of orbits were computed, using different staring distances from the center of the galaxy, with the same low values of the initial radial and vertical velocities as was mentioned before. Orbits starting at the same distance $r$, are considered to belong to the same group. We have chosen this method of approach because the alternative, a self-consistent treatment with the energy as an additional parameter, in a scattering setting and in a galactic potential, is beyond the scope of the present paper. Some of these ideas may indeed be incorporated into a future work. Our numerical results suggest that, for each group of orbits, there is a linear relationship between $L_{zc}$ and $M_n$, similar to that shown in Fig. 3.

It is also of interest to compare the critical angular momentum $L_{zc}$ with the circular angular momentum $L_{z0}$. The circular angular momentum is given by the equation
\begin{equation}
L_{z0} = \left[r_0^3 \left(\frac{\partial \Phi_t}{\partial r}\right)_{r=r_0,z=0}\right]^{1/2} \ \ \ .
\end{equation}
Such a comparison, for five values of the mass of the nucleus, when $c_n=0.25$ kpc, is given in Table 1. Orbits start at three different distances from the galactic center and according to the above, belong to three groups of orbits. It is important to observe, that the ratio $L_{zc}/L_{z0}$ increases with decreasing $r$. This means that more stars (initially in the plane) are leaving the galactic plane resulting from chaos in the central region, than for larger values of $r$. This phenomenon could be taken as a first step for the explanation of the existence of a chaotic bulge in most disk-shaped galaxies, or perhaps, the presence of chaotic orbits ``contaminating" some more conventional stellar dynamical bulge.
\begin{table}[ht]
\caption{A comparison between $L_{zc}$ and $L_{z0}$ for different distances from the galactic center.}
\centering
\setlength{\tabcolsep}{4.0pt}
\begin{tabular}{|c||c c c c c c c c c|}
\hline \hline
 & \multicolumn{3}{c}{$r=2.5$} & \multicolumn{3}{c}{$r=5.0$} & \multicolumn{3}{c}{$r=8.5$} \bigstrut \\ \hline
$M_n$ & $L_{zc}$ & $L_{z0}$ & $L_{zc}/L_{z0}$ & $L_{zc}$ & $L_{z0}$ & $L_{zc}/L_{z0}$ & $L_{zc}$ & $L_{z0}$ & $L_{zc}/L_{z0}$ \\
\hline \hline
100 & 15 & 30 & 0.516 & 27 & 90 & 0.300 & 28 & 190 & 0.147 \\
200 & 19 & 34 & 0.558 & 36 & 93 & 0.388 & 37 & 192 & 0.191 \\
300 & 24 & 37 & 0.648 & 45 & 95 & 0.472 & 46 & 195 & 0.236 \\
400 & 28 & 40 & 0.693 & 53 & 98 & 0.546 & 55 & 197 & 0.279 \\
500 & 32 & 43 & 0.739 & 62 & 100 & 0.617 & 64 & 200 & 0.321 \\
\hline \hline
\end{tabular}
\end{table}

\section{Evolution of orbits in the time dependent model}

In this Section, we shall study the evolution of orbits in our galactic model, when mass transport is taking place. We assume that mass is transported from the disk to the nucleus, in such as way, that we have an exponential increase in the mass of the nucleus, while an exponential decrease occurs in the mass of the disk. The exponential mass transportation follows the equations
\begin{eqnarray}
M_{df} &=& M_{di} - m \left(1-e^{-kt}\right), \nonumber \\
M_{nf} &=& M_{ni} + m \left(1-e^{-kt}\right),
\end{eqnarray}
where $M_{di}$ and $M_{ni}$ are the initial values of the mass of the disk and the nucleus respectively, $m$ is the portion of the disk mass that is transferred, $t$ is the time, while $k>0$ is a positive parameter, which determines the timescale of the evolution. This particular model of mass transportation was introduced in Caranicolas \& Papadopoulos (2003). In our study, we will investigate in more detail this model. In particular, we will try to find out the correlation, if any, between the critical value of the angular momentum $L_{zc}$ and the final mass of the nucleus $M_{nf}$. As final mass we consider the limit of the current mass $M_{nf}$, as $t \rightarrow \infty$.
\begin{figure}[!tH]
\centering
\resizebox{0.95\hsize}{!}{\rotatebox{0}{\includegraphics*{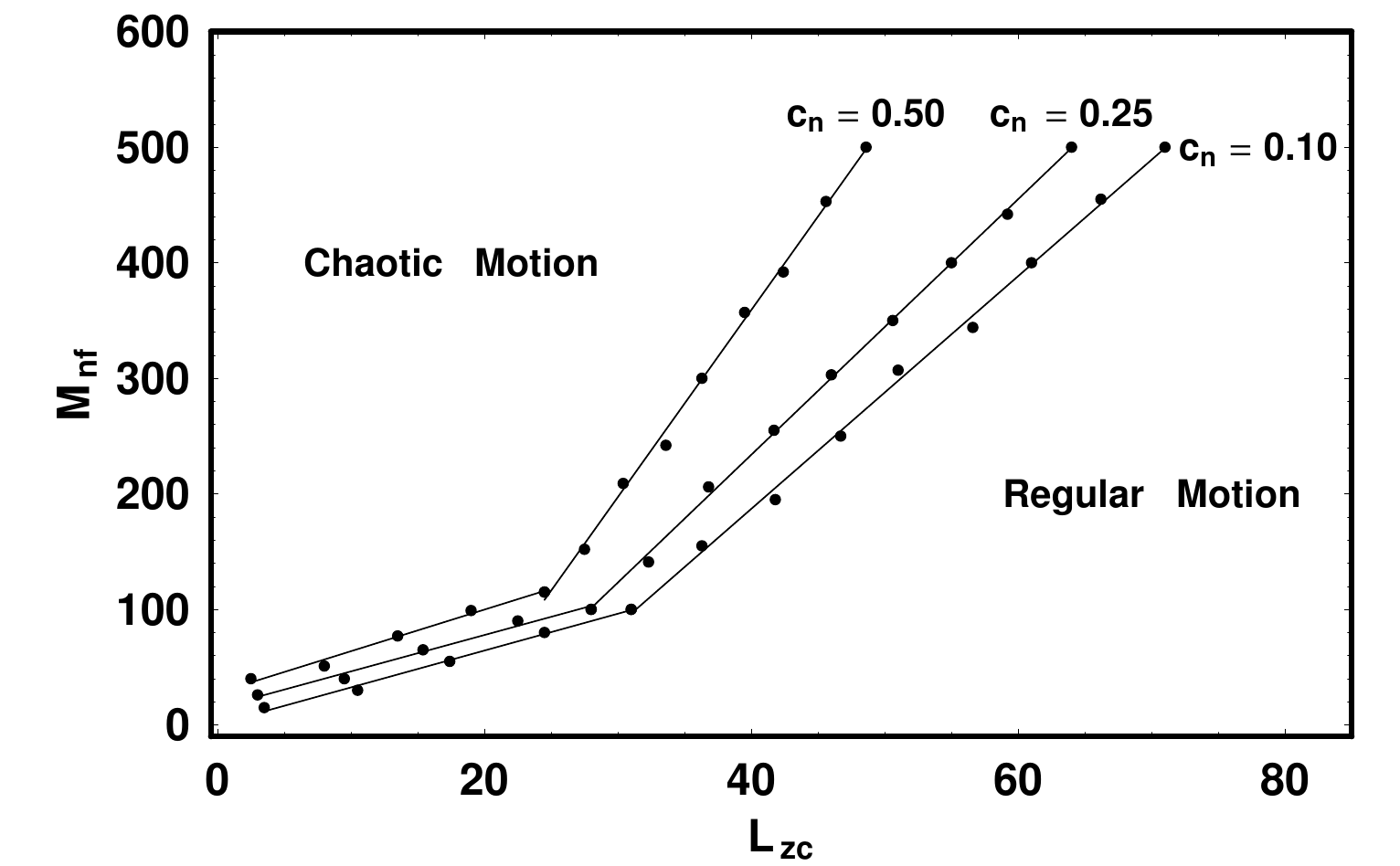}}}
\caption{Relationship between the critical value of the angular momentum $L_{zc}$ and the final mass of the nucleus $M_{nf}$. Details are given in the text.}
\end{figure}
\begin{figure*}[!tH]
\centering
\resizebox{0.9\hsize}{!}{\rotatebox{270}{\includegraphics*{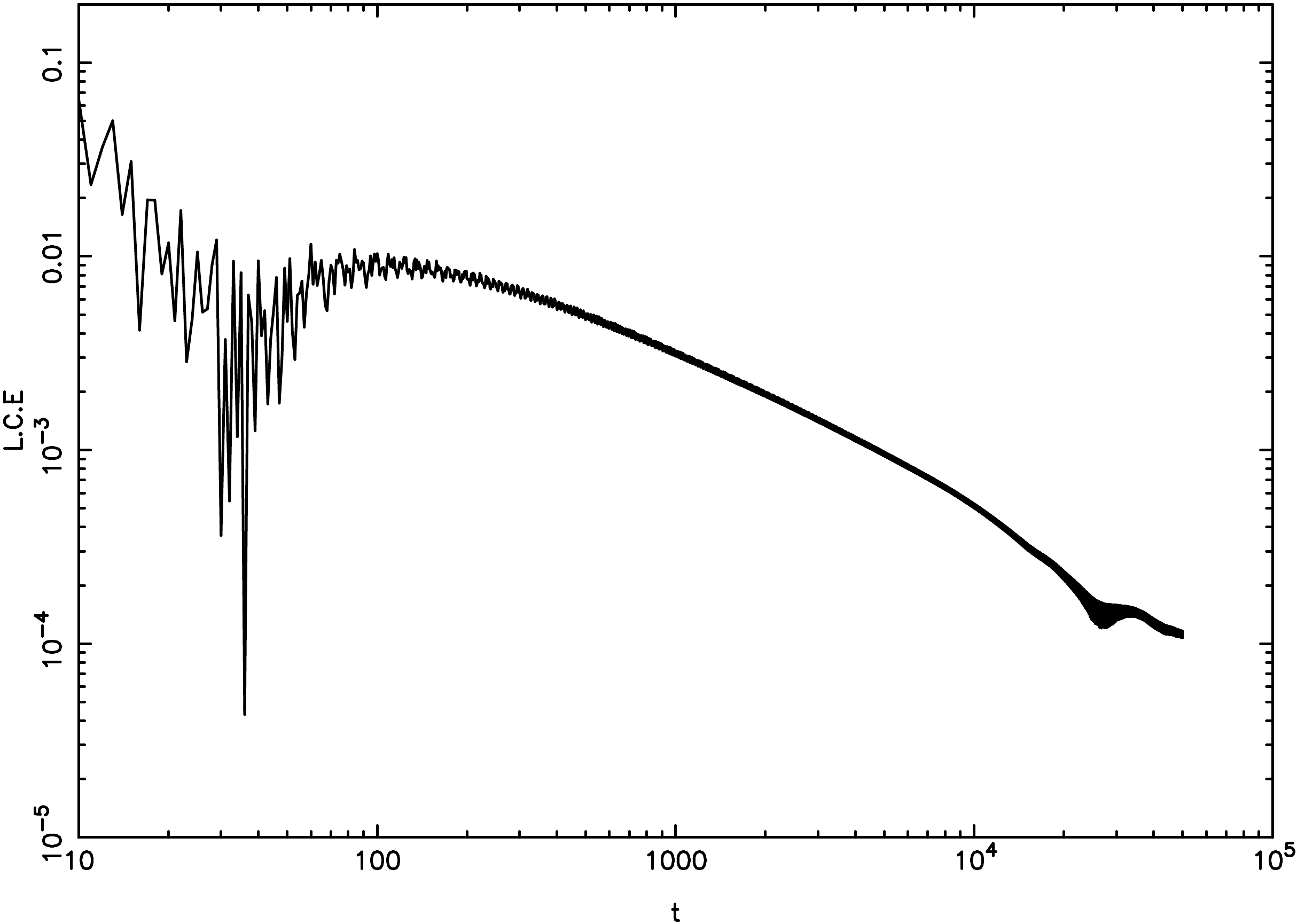}}\hspace{2cm}
                         \rotatebox{270}{\includegraphics*{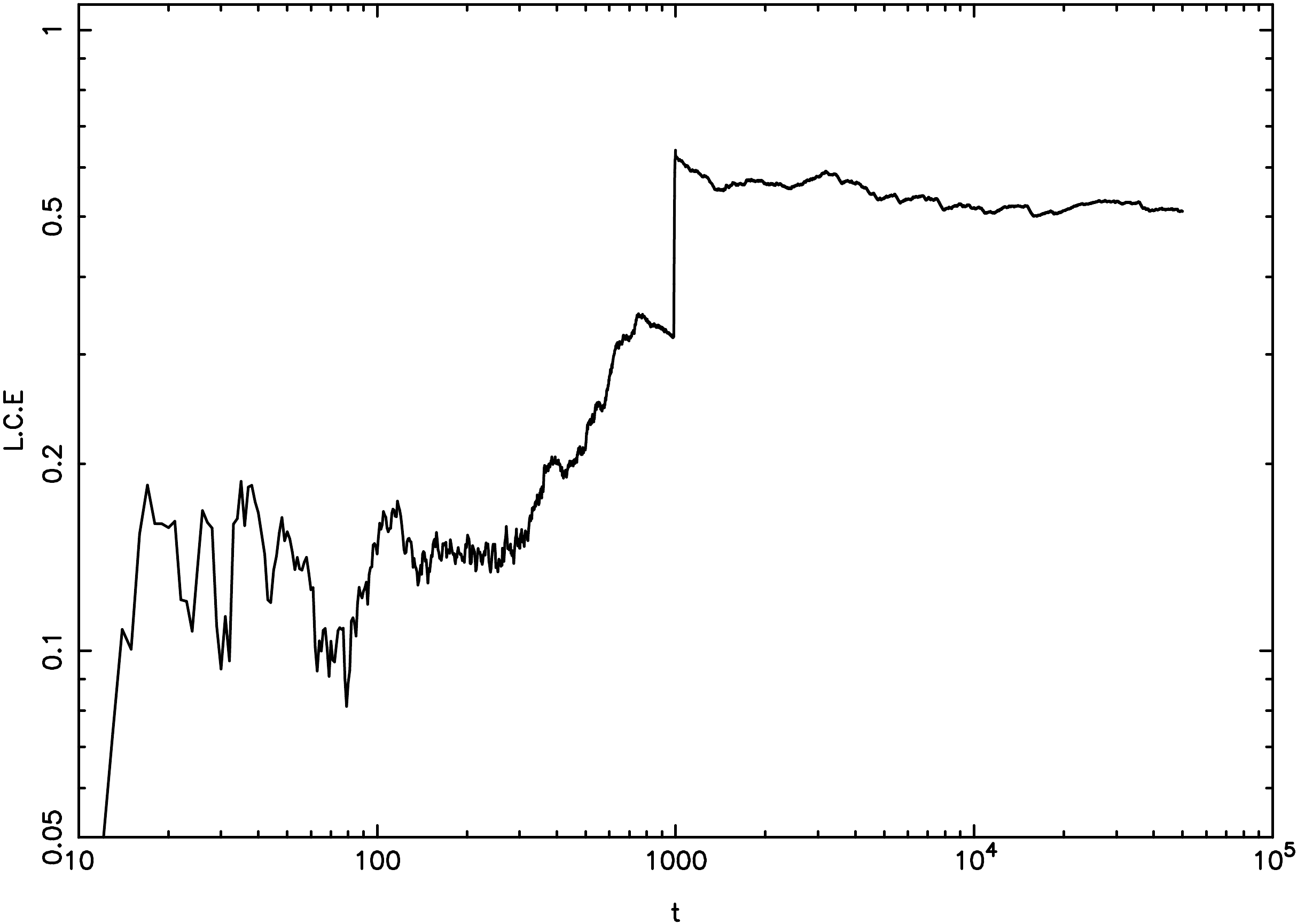}}}
\resizebox{0.9\hsize}{!}{\rotatebox{270}{\includegraphics*{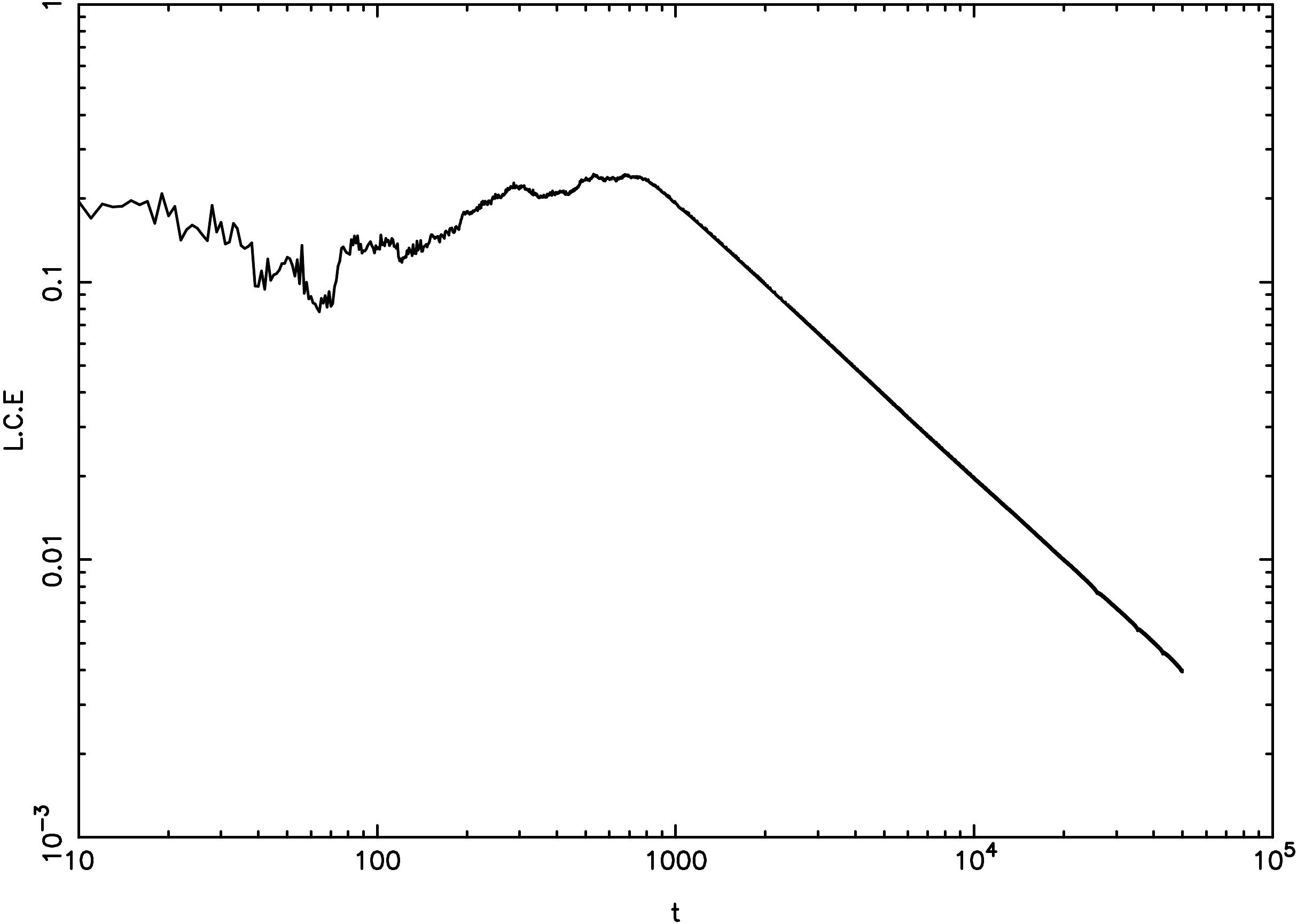}}\hspace{2cm}
                         \rotatebox{270}{\includegraphics*{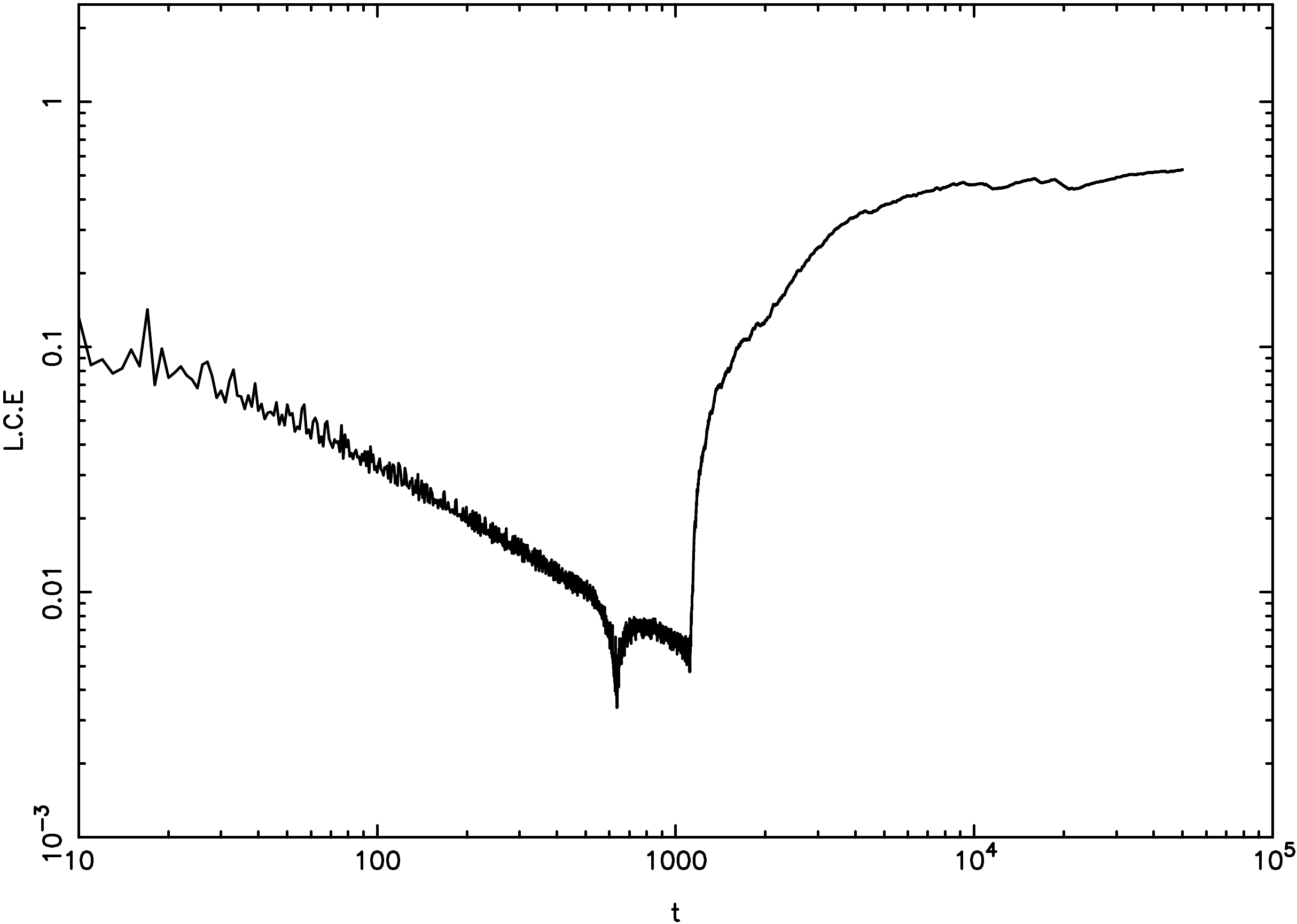}}}
\vskip 0.1cm
\caption{(a-d): Evolution of the L.C.E of four different orbits, following the set of Eqs. (20). By this procedure a massive nucleus is formed in the core of the galaxy. (a-upper left): The orbit starts as a regular and remains regular, (b-upper right): the orbit starts as chaotic and remains chaotic, (c-down left): an orbit which starts as chaotic but after 1000 time units, when the evolution stops it becomes regular and (d-down right): an orbit starting as regular but after the mass transportation it changes its nature to chaotic. Details are given in the text.}
\end{figure*}

Orbits were started at $r_0=8.5, z_0=0$ with radial and vertical velocities smaller than 30 km/s, while $k=0.004$ and $m=300$. Due to the fact that the corresponding Hamiltonian is time dependent, it is not possible to use again the Poincar\'{e} surface of section technique. It is well known, that the shape of orbits, sometimes is inconclusive or misleading. In order to overcome this drawback, we decided to use a highly accurate method, such as the Lyapunov Characteristic Exponent - L.C.E. The main advantage of this method is that it uses certain and objective numerical thresholds beyond which we can distinguish between ordered and chaotic motion. On the other hand, this method is very time consuming, as it needs time intervals of numerical integration of order of $10^5$ time units, in order to give reliable and definitive results regarding the nature of an orbit. But this is a ``price" we can afford to pay.
\begin{figure*}[!tH]
\centering
\resizebox{\hsize}{!}{\rotatebox{0}{\includegraphics*{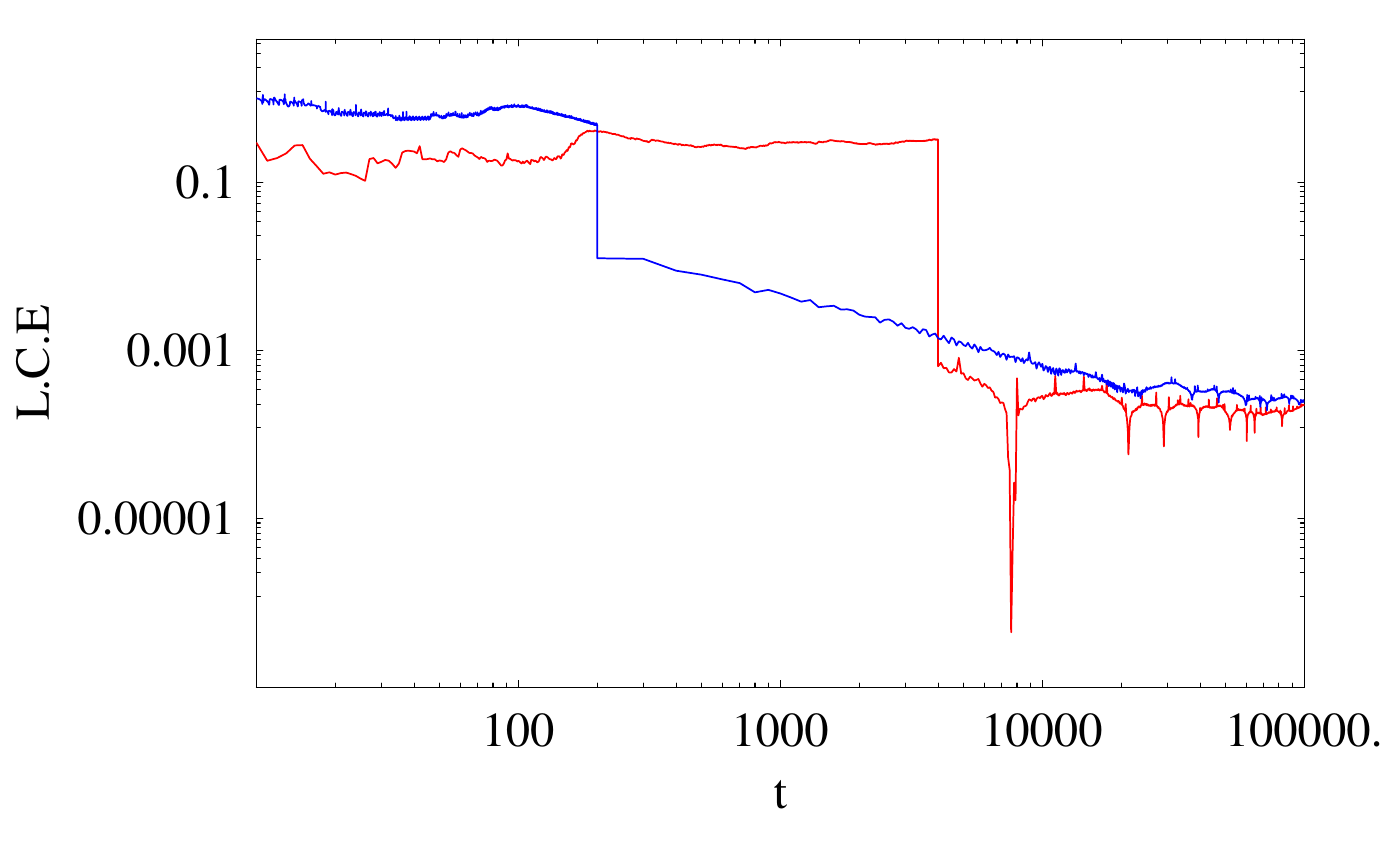}}\hspace{2cm}
                      \rotatebox{0}{\includegraphics*{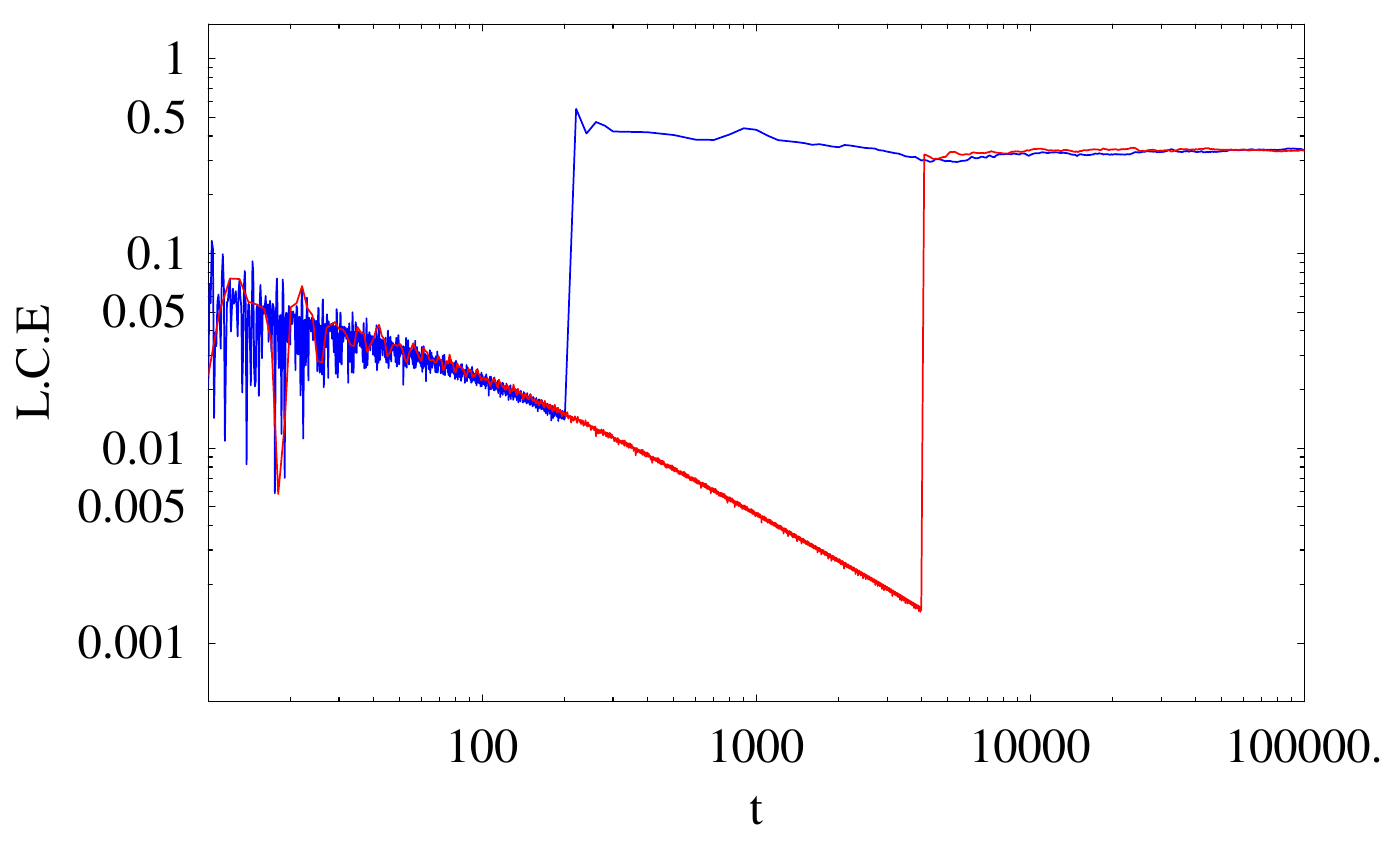}}}
\vskip 0.1cm
\caption{(a-b): Evolution of the L.C.E of the same orbits, for different values of the timescale $k$. (a-left): The orbit starts as chaotic but after the mass transportation it becomes regular and (b-right): the orbit starts as regular but after the mass transportation it alters its nature to chaotic. The blue line stand for $k=0.04$, while the red one for $k=0.001$. Details are given in the text.}
\end{figure*}

Our numerical calculations suggest that a linear relationship exists between $L_{zc}$ and $M_{nf}$. This can be easily seen looking at the diagram shown in Figure 13. Each line divides the $\left[L_{zc} - M_{nf}\right]$ plane in two parts. Orbits with values of the physical parameters $L_{zc}$ and $M_{nf}$ on the upper left hand side of the diagram, including the line, are chaotic, while orbits with initial conditions on the lower right hand side part of the same diagram are regular. Note that, for very small values of the angular momentum, the slope of the lines is different. In order to produce this part of the diagram, we used small values of $M_{ni}$ and $m$. In all other cases we have used $M_{ni} > 100$. One observes that this diagram is very similar to the diagram given in Fig. 3, which corresponds to the time independent model. Moreover, numerical calculations, not provided here, suggest that if we change the value of the timescale factor $k$, the outcomes would be very similar again. Therefore, we may conclude that the mass transportation (itself or the time scale factor $k$) does not affect so much the correlation between $L_{zc}$ and $M_n$.
\begin{table}[ht]
\caption{The respective percentage of total tested orbits for different values of the timescale $k$.}
\centering
\setlength{\tabcolsep}{6pt}
\begin{tabular}{|c|| c | c | c | c|}
\hline \hline
$k$ & $R \rightarrow C$ & $C \rightarrow C$ & $R \rightarrow R$ & $C \rightarrow R$ \\
\hline \hline
0.001 & 56\% & 25\% & 16\% & 3\% \\
0.004 & 58\% & 24\% & 14\% & 4\% \\
0.04 & 61\% & 26\% & 11.8\% & 1.2\% \\
\hline \hline
\end{tabular}
\end{table}

Figure 14a-d shows the evolution of the Lyapunov Characteristic Exponent of four different orbits, as the total mass distribution of the dynamical system changes with time, following the set of equations (20). For all orbits shown in Fig. 14a-d, the initial value of the Hamiltonian (8) is $E_i=-950$, while $M_{ni}=100, c_n=0.25, m=300$ and $k=0.004$. In Figure 14a we can see the evolution of the Lyapunov Characteristic Exponent for an orbit and for a time period of $10^5$ time units, as the galaxy evolves. The initial conditions are: $r_0=2.2, z_0=0, p_{r0}=0$, while the value of $p_{z0}$ is found from the Hamiltonian (8) in all cases. The value of the angular momentum is $L_z=1$. When $t=1000$ time units, the mass of the nucleus is $M_{nf}=400$ and the evolution stops. The value of the Hamiltonian is now $E_f=-958$. Note that the L.C.E tends asymptotically to zero, which clearly indicates that this orbit starts as regular and remains regular always near the galactic plane, during the galactic evolution. Figure 14b is similar to Fig. 14a. This orbit has initial conditions: $r_0=8.65, z_0=0, p_{r0}=0$, while the value of the angular momentum is $L_z=10$. When $t=1000$ time units, the final mass of the nucleus is $M_{nf}=400$ and the mass transportation stops. The value of the Hamiltonian is now $E_f=-961$. In this case, the profile of the L.C.E clearly indicates that this orbit starts as a chaotic orbit and remains chaotic, during the galactic evolution. We can observe that as the mass of the nucleus increases, during the mass transportation, the value of the L.C.E also increases gradually. Figure 14c depicts the evolution of the L.C.E of an orbit with initial conditions: $r_0=8.47, z_0=0, p_{r0}=0$. The value of the angular momentum is now $L_z=20$. After a time interval of 1000 time units the mass of the nucleus is $M_{nf}=400$ and the galactic evolution stops. The Hamiltonian settled to the value $E_f=-959$. The profile of the L.C.E given in Fig. 14c indicates that this orbit stars as a chaotic orbit but after the mass transportation it becomes an ordered one. In Figure 14d we observe the evolution of the L.C.E of an orbit with initial conditions: $r_0=8.8, z_0=0, p_{r0}=0$, while the value of the angular momentum this time is $L_z=30$ . After a time interval of 1000 time units the final value of the mass of the nucleus is $M_{nf}=400$ and therefore, galactic evolution stops. The Hamiltonian settled to the value $E_f=-960$. The profile of the L.C.E given in Fig. 14d shows that this orbit stars as a regular orbit but after the mass transportation, it becomes a chaotic one. Therefore, it is evident that if mass transport was not present, the orbit this time would have remain regular. In this case, the presence of the massive nucleus in the core of galaxy, has changed the nature of the orbit, from regular to chaotic. Furthermore, one can say that we are dealing with the case of ``fast" chaos, because the corresponding value of the L.C.E is relatively large (see for details in Caranicolas, 1990 and references therein).

As one would expect, the orbital evolution of stars in an evolving galactic model and the transition from regularity to chaos, should strongly depends on the timescale of the evolution, determined by the $k$ parameter. In order to reveal the importance of the timescale, we will explore several cases, in which the temporal evolution timescale is varied. Figure 15a shows the evolution of the L.C.E of the orbit with the same initial conditions as in Fig. 14c, for two different values of the timescale $k$. The blue line stand for $k=0.04$, while the red one for $k=0.001$. In both cases, the orbit starts as a chaotic but after the mass transportation ($t > 200$ time units for the orbit shown in blue line and $t > 4000$ time units for the orbit shown in red line) the orbit becomes regular. Note that both lines seem to converge when $t > 10^{4}$ time units. Similar in Figure 15b, we can see the the evolution of the L.C.E of the orbit with the same initial conditions as in Fig. 14d, for two different values of the timescale $k$. Again, the blue line stand for $k=0.04$, while the red one for $k=0.001$. In both cases, the orbit starts as a regular but after the mass transportation ($t > 200$ time units for the orbit shown in blue line and $t > 4000$ time units for the orbit shown in red line) the orbit alters its nature and becomes chaotic. Note once more, that the values of the L.C.E converge when $t > 4000$ time units. One should be very careful because, for small time intervals, a regular orbit may look chaotic and vise versa. This is due to the different values of the timescale $k$, which determines how fast the mass from the disk is transported to the nucleus. This phenomenon is observed only when the galactic evolution (mass transportation) takes place. When the galactic evolution stops and the system becomes time independent, the nature of the orbits is the same, regardless the particular value of $k$ we used. Thus, we reach to the conclusion that the timescale itself, does not affect directly the nature of the orbits during the galactic evolution. It only determines each time, the time interval which is needed for an orbit during the mass transportation in order to form its final character (regular or chaotic).

Using the above procedure, we have computed and tested a large number of orbits (approximately 3000 different initial conditions of orbits for three different values of the timescale $k$), with different starting distances from the center of the galaxy, while keeping the corresponding radial and vertical velocities smaller than 30 km/s. Orbits starting at the same distance $r$ were considered to belong to the same group. All numerical outcomes suggest that, for each group of orbits, a linear relationship exists, between the critical value of the angular momentum $L_{zc}$ and the final mass of the nucleus $M_{nf}$. Our numerical results indicate that the character of the orbits can change either from regular to chaotic and vise versa or not change at all, as the mass is transported and the massive nucleus is developed in the central region of the disk galaxy. Table 2 shows the respective percentage of total tested orbits in each case. The symbols $R \rightarrow C$, $C \rightarrow C$, $R \rightarrow R$ and $C \rightarrow R$ correspond to each kind of the alteration of the orbital nature (i.e. $R \rightarrow C$ corresponds to Regular to Chaotic etc). From Table 2, it is evident that the formation of a massive nucleus, leads the majority of the orbits of the system to alter their nature from regular to chaotic or to stay chaotic. Some orbits maintain their regular character but only a very small portion of the total tested orbits change their nature from chaotic to regular during the mass transportation. This tiny portion of initial conditions belongs to resonant orbits of higher multiplicity which correspond to multiple small islands of invariant curves in the $(r,p_r)$ phase plane. These resonant orbits appear when the nucleus reaches its final value, that is $M_{nf}=400$. Furthermore, we may conclude that the particular value of the timescale $k$ does not affect directly the galactic evolution. The dominant parameter which determines the nature of an orbit is the final mass of the nucleus. The timescale $k$, simply shows how fast the galactic nucleus would reach its final mass through the mass transportation. Whether the nature of an orbit will change or not during the galactic evolution described by the set of equations (20), strongly depends on the initial conditions $\left(r_0,p_{r0}\right)$ of each orbit. Moreover, as the change of the value of the Hamiltonian (8) in each case is practically negligible, $\left(\Delta E \lesssim 1.5 \%\right)$, we can say that the orbits in each group can be considered as isoenergetic.

\section{Discussion and conclusions}

In the present paper, we used an axially symmetric galactic gravitational model with a disk-halo and a spherical nucleus, in order to investigate the transition from regular to chaotic motion for stars moving in the meridian $(r,z)$ plane. We studied in detail, the transition from ordered to chaotic motion, in two different cases: the time independent model and the time evolving model, when mass is transported from the disk to central nuclear region. In both cases, we explored all the available range regarding the values of the main involved parameters of the dynamical system and we obtained interesting results, regarding the correlations between the important physical quantities of the system.

Here, we have to point out which of the outcomes of the present study are novel. In this research we have used a composite disk-halo potential with an additional spherically symmetric nucleus, represented by a Plummer sphere. The current paper is basically consists of two parts. In the first part which covers Sections 3 and 4, describes the time independent galactic model. He have conducted a thorough, detailed and systematic study in order to connect the involved parameters of the dynamical system. In particular we obtained the relationship which connects the critical value of the angular momentum with the mass of the nucleus. Moreover, we connected the mass of the nucleus with the chaotic percentage and also with the average value of the Lyapunov Characteristic Exponent, for several values if the angular momentum. All these numerically obtained relationships, have been reproduced and verified using a semi-theoretical approach. The second part of the present paper, that is Section 5, describes the time dependent model. Here, we adopted the simple model of mass transportation proposed by Caranicolas \& Papadopoulos (2003), as a starting point. We studied in much detail this model using three different values of the timescale $k$ of the galactic evolution. For each value of the $k$ parameter, we computed 1000 different and random initial conditions $\left(r_0, p_{r0}\right)$. Our calculations suggest that the majority of the orbits of the system alter their nature from regular to chaotic or remain chaotic. Some orbits maintain their regular character but only a very small portion of the total tested orbits change their nature from chaotic to regular during the mass transportation. It was found, that the particular value of the timescale $k$ does not affect directly the galactic evolution. It is only determines how fast the nucleus would reach its final value of the mass. The most significant revelation is that the dominant parameter, which determines whether an orbit would change or not and how its orbital nature during the mass transportation, is the final value of the mass of the nucleus.

It is evident that there is a well defined transition from regularity to chaos in our galactic model, for a given mass of the disk-halo component. This transition is nearly a straight line connecting the points in the $\left[L_{zc} - M_n \right]$ plane. The slope of this line increases with the scale length of the nucleus $c_n$. The physical interpretation of this result is that, a nucleus of the same mass but with greater concentration (smaller $c_n$), deflects stars with higher values of angular momentum. It also evident that the above straight line relation can be derived and reproduced using elementary mathematical calculations. For this derivation, we have assumed that the deflection into a vertical chaotic orbit, happens after $n$ encounters with correlated deflections. This choice was made by observing the behavior of the orbits in our numerical simulations and also by taking into account the smoothness of the total $F_z$ force (see Fig. 11).

In order to keep things simple, we have kept some of the parameters of the system constant and studied the behavior of orbits varying only three basic parameters $\left(M_n, c_n, L_z \right)$. It was found that, there is a linear relationship between the average value of the L.C.E and the mass of the nucleus $M_n$. On the contrary, the relation between the chaotic percentage $A\%$ on the $(r,p_r)$ phase plane and the mass of the nucleus $M_n$ is not linear but rather exponential. The numerically found outcomes were explained, using some semi-theoretical arguments. Moreover, we concluded that when the scale length of the nucleus $c_n$ is smaller, that is when the nucleus is more dense, the $F_{zn}$ force is stronger, for each particular value of the mass of the nucleus $M_n$ and thus the chaoticity of the dynamical system increases.

A comparison of the critical angular momentum $L_{zc}$ to the local circular angular momentum $L_{z0}$, reveals that the ratio $L_{zc}/L_{z0}$ increases as we approach the center of the galaxy. For a galaxy with a nearly flat rotation curve, this variation mainly reflects the variation of the galactocentric distance $r$. Therefore, stars moving near the center can more easily be deflected to higher values of $z$. This could be connected with the presence of massive spherical nuclei in the central regions of disk galaxies and suggest that near such massive nuclei the nature of the orbits of the stars must be highly chaotic, i.e., with isotropic velocity dispersions. Our numerical experiments, show that near the center, the total $F_z$ force is entirely due to the massive nucleus and behaves like that of a harmonic oscillator. This is a strong indication, that under certain conditions, the harmonic oscillator approximation can provide reliable results regarding the motion in the central regions of real galaxies.

Today, we know that the addition of matter onto compact objects, such as massive and dense galactic nuclei, is a very complicated astrophysical mechanism, which it would be very difficult to be studied and explored, if we do not establish some basic assumptions in order to make it more maneuverable. In the present research, we have studied the change in the nature of the orbits in a time-dependent model, with an exponential mass transport from the disk to the nuclear region. We have adopted this simple model, without going into more details for two basic reasons. The first reason is that as it was mentioned above, it would be very difficult to construct a model in order to explain such a complicated mechanism and the second reason is that we are not interested in the astrophysical structure of the galaxy, but only in its dynamical behavior. Here we have to clarify, that our simple time dependent model, may not necessarily reflect the actual astrophysical mechanism. Nevertheless, similar simple dynamical models have been used frequently in the past (see Caranicolas \& Papadopoulos, 2003; Caranicolas \& Innanen, 2009) in order to study the dynamical behavior of stellar orbits in complicated galactic systems. Once more, we have to point out that our current treatment of the system is purely dynamical. This is exactly the main scope of this article. It is well known that such mechanisms are met in the central regions of active galaxies and they are responsible for the enormous luminosity of the quasars which are hosted in the cores of the galaxies (see Collin \& Zahn, 1999 and references therein).

Our numerical calculations show that a linear relationship between the the critical value of the angular momentum $L_{zc}$ and the final mass of the nucleus $M_{nf}$ also exists. Note, that the slope of this line increases as the value of the scale length of the nucleus $c_n$ increases. This fact indicates that a nucleus of the same mass but with larger concentration (smaller value of the $c_n$) is able to deflect stars with greater values of the angular momentum. Therefore, one can say that the present results are similar to those obtained in Zotos (2011b), where we have also connected the involved parameters of the system with the extent of the chaotic regions and the angular momentum. In addition to the observed similarities, we have also observed some differences in the behavior of orbits, between this model and the model used by Zotos (2011a). The most important difference is that in our previous work, the stars moved towards the halo after several encounters with the nucleus. Here stars can stay for long time periods (this actually depends on the particular value of the timescale $k$, that is on how fast mass is transported from the disk to the nucleus), before they scattered to the halo. Therefore, one must be very cautious because, for small time periods, a chaotic orbit may look regular. It seems that, in this case, one is not able to use methods for fast detection of chaos, especially when small values of $k$ are used (see Karanis \& Caranicolas, 2002 and references therein).

Going on step further into the physical interpretation of the outcomes of the present work, we can summarize the following: There is no doubt that a large number of galaxies show significant activity in their nuclear regions. We adopted a simple model, in order to express this activity, which is described by a mass transportation from the disk to the nuclear region. The novelty and the aim of this research, is to explore on how the dynamical properties of the galaxy, that is the behavior of orbits, have been affected by the mass transfer phenomenon and also to establish a relationship between two important physical quantities: the final mass of the nucleus and the star's angular momentum. The physical conclusion is that in active galaxies, a large number of low angular momentum stars display chaotic motion. Furthermore, some orbits that may look regular, in fact must be chaotic when the transportation of the mass is slow. The general conclusion from our work is that the dominant parameters, which determine the nature of the orbits in galaxies are the mass of the nucleus and the angular momentum. The rest involved quantities do not affect directly the nature of the orbits. In particular, as we have seen in the time dependent model, the timescale $k$ does not really affect the final character of the orbits. It is in our future plans to try a different way of mass transportation (i.e a linear). From the present outcomes we may foresee, until proven, that even a different kind of mass transportation would not affect directly the nature of the orbits, since as we have seen the dominant parameter is the final mass of the nucleus and not how (how fast or with which way) we reach this value.

\section*{Acknowledgments}

\textit{I am much indebted to Dr. Hagai Perets for reviewing the manuscript and also for all his very useful suggestions and comments which improved significantly the quality and the clarity of the present work.}

\end{document}